\documentstyle[prc,aps,preprint,epsf]{revtex}

\draft
\tighten
\preprint{GSI-Preprint-98-15}

\begin{document}

\title{
Breakup Conditions of Projectile Spectators from Dynamical Observables
}

\author
{
M.~Begemann-Blaich$^{1}$,
V.~Lindenstruth$^{1,9}$,
J.~Pochodzalla$^{1,2}$,
J.C.~Adloff$^{3}$,
P.~Bouissou$^{4}$,
J.~Hubele$^{1}$,
G.~Imme$^{5}$,
I.~Iori$^{6}$,
P.~Kreutz$^{7}$,
G.J.~Kunde$^{1,10}$,
S.~Leray$^{4}$,
Z.~Liu$^{1,11}$,
U.~Lynen$^{1}$,
R.J.~Mei$\!$jer$^{1}$,
U.~Milkau$^{1}$,
A.~Moroni$^{6}$,
W.F.J.~M\"{u}ller$^{1}$,
C.~Ng\^{o}$^{4}$,
C.A.~Ogilvie$^{1,12}$,
G.~Raciti$^{5}$,
G.~Rudolf$^{3}$,
H.~Sann$^{1}$,
M.~Schnittker$^{1}$,
A.~Sch\"uttauf$^{2}$,
W.~Seidel$^{8}$,
L.~Stuttge$^{3}$,
W.~Trautmann$^{1}$,
A.~Tucholski$^{1}$
}

\address
{
$^{1}$ Gesellschaft f\"ur Schwerionenforschung, D-64220 Darmstadt, Germany
\\
$^{2}$ MPI f\"ur Kernphysik Heidelberg, D-69029 Heidelberg, Germany
\\
$^{3}$ Centre de Recherches Nucl\'{e}aires, F-67037 Strasbourg, France
\\
$^{4}$ Laboratoire National Saturne, CEN Saclay, F-91191 Gif-sur-Yvette, France
\\
$^{5}$ Dipartimento di Fisica dell' Universit\'{a} and I.N.F.N., 
       I-95129 Catania, Italy
\\
$^{6}$ Istituto di Scienze Fisiche dell' Universit\'{a}
       and I.N.F.N., I-20133 Milano, Italy
\\
$^{7}$ Institut f\"ur Kernphysik,
       Universit\"at Frankfurt, D-60486 Frankfurt, Germany
\\
$^{8}$ Forschungszentrum Rossendorf, D-01314 Dresden, Germany
\\
$^{9}$ Nuclear Science Division, LBL, Berkeley, CA 94720, USA
\\
$^{10}$ Dept.~of Physics, Yale University, New Haven, CT 06512, USA
\\
$^{11}$ IHEP, Beijing 100039, P.R.~China
\\
$^{12}$ Department of Physics, MIT, Cambridge, MA 02139, USA
}

\date{\today}

\maketitle

\begin{abstract}
Momenta and masses of heavy projectile fragments ($Z \geq 8$),
produced in collisions of $^{197}$Au with C, Al, Cu and Pb targets
at E/A = 600 MeV, were determined with the ALADIN magnetic spectrometer
at SIS. Using these informations, an analysis of kinematic correlations between
the two and three heaviest projectile fragments in their rest frame was
performed. The sensitivity of these correlations to the
conditions at breakup was verified within
the schematic SOS-model. For a quantitative investigation,
the data were compared to calculations with statistical multifragmentation
models and to classical three-body calculations.

With classical trajectory calculations, where the charges and masses of the
fragments are taken from a Monte Carlo sampling of the experimental events,
the dynamical observables can be reproduced.
The deduced breakup parameters, however, differ considerably from those
assumed in the statistical multifragmentation models which 
describe the charge correlations. If, on the other hand,
the analysis of kinematic and charge correlations is performed for events
with two and three heavy fragments produced by statistical multifragmentation
codes, a good agreement with the data is found with the exception that
the fluctuation widths of the intrinsic fragment energies are significantly
underestimated.
A new version of the multifragmentation code MCFRAG was therefore used to
investigate the potential role of angular momentum at the breakup stage.
If a mean angular momentum of 0.75$\hbar$/nucleon is added to the system, the
energy fluctuations can be reproduced, but at the same time the charge
partitions are modified and deviate from the data.
\end{abstract}

\pacs{PACS number(s):
  25.70.Mn,
  25.70.Pq, 
  25.75.Ld, 
  25.75.-q
}

\narrowtext

\section{Introduction}
\label{SEC1}

%\noindent
In several experiments with the ALADIN spectrometer, the decay of excited
projectile spectator matter at beam energies between 400 and 1000 MeV per
nucleon
was studied \cite{HUB91,KRE93,SCH96}. In these collisions, energy depositions
are reached which cover the range from particle evaporation to
multi-fragment emission and further to the total disassembly of
the nuclear matter, the so-called 'rise and fall of multifragment emission'
\cite{OGI92}.
The most prominent feature of the multi-fragment decay is
the universality that is obeyed by the fragment multiplicities and the
fragment charge correlations. These observables are invariant with
respect to the entrance channel --~i.e.~independent of the beam energy
and the target~-- if plotted as a function of $Z_{bound}$, where $Z_{bound}$
is the sum of the atomic numbers $Z_{i}$ of all projectile fragments with
$Z_{i} \geq 2$. For different projectiles, the dependence of the fragment 
multiplicity on $Z_{bound}$ follows a linear scaling law.
These observations indicate that
compressional effects are only of minor importance. In contrast
to central collisions at lower energies, where large radial flow
effects are observed, the quantitative interpretation of
kinematic observables is therefore simplified.

%\noindent
More important,
these characteristics are an indication that chemical equilibrium
is attained prior to the fragmentation stages of the reaction. In fact,
statistical models were found to be quite successful in describing the 
experimental
fragment yields and charge correlations, if the breakup of an expanded system
was assumed \cite{BAO93,BAR93,BOT92,BOT95,BOT85,BON85,GRO86}.
In addition, the temperature of the excited matter, extracted
from double ratios of isotope yields, is reproduced. 
On the other hand, the kinetic energy spectra of particles
and fragments are not equally well described within the statistical picture.
The energy spectra of light charged particles ($A \leq 4$) can be explained
by a thermal emission of the fragments,
but their slopes correspond to temperatures approximately three times larger
than those extracted  from isotope ratios \cite{XI97}. While this may be an
indication for pre-breakup emission if remains to be understood whether the
kinetic energies of intermediate mass fragments ($ 3 \leq Z \leq 30$) are
consistent with the statistical approach.

%\noindent
The dynamics of the multifragmentation process has therefore to be studied. 
It is well known that kinematic
correlations, which are governed by the long range Coulomb repulsion, are 
sensitive
to the disintegration process. Previous studies concentrated mostly on the
two-fragment velocity correlation functions
\cite{TRO87,GRO89,KIM91,KIM92,BOW93,BAU93,KAE93,BAO94,SCH94}.
Only few attempts were made to analyze higher order correlations. However, these
studies were done either for heavy fragments at much lower beam energies
\cite{GLA83,PEL86,BOU89,BIZ92,BRU94} or for light charged particles only 
\cite{LAU94}.
In this paper, the results of a kinematic analysis of the fragmentation process
of the projectile spectator are presented. Heavy projectile
fragments produced in peripheral Au induced collisions at
E/A = 600 MeV are studied without the influence of energy thresholds 
of the detectors. Moreover, the analysis is performed in the center of mass
frame of the fragments, thus reducing the influence of directed collective
motion of the emitting source. On the other hand, a limit of $Z \geq 8$ is
imposed for kinematic observables by the lower detection threshold of the 
TP-MUSIC II
tracking detector. The analysis is therefore restricted 
to the excitation energy range characterized by
increasing fragment multiplicities.

\section{The experiment}
\label{SEC2}

\subsection{Experimental setup}
\label{SEC21}

\begin{figure}
      \epsfxsize=8.0cm
      \centerline{\epsffile[40 0 410 230]{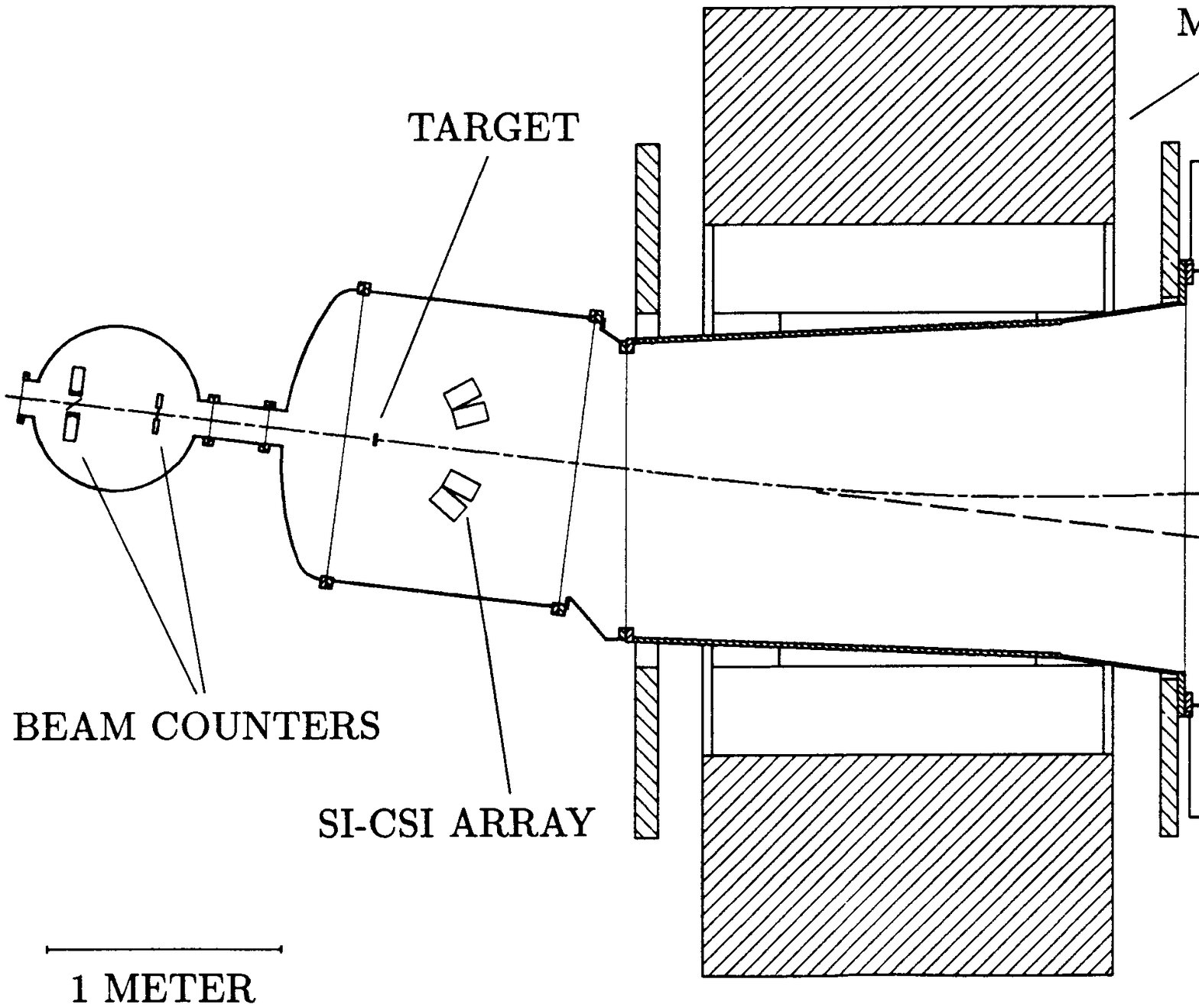}}
 \caption [] {\protect
  Schematic view of the experimental setup in the bending plane of the magnet.
  The beam enters from the left and hits the beam counters before reaching the
  target. Mid-rapidity particles are detected in the Si-CsI array. 
  Projectile fragments are tracked and identified in the TP-MUSIC II detector
  and in the time-of-flight wall.
 }
 \label{setup}
 \vspace{0.1cm}
 \end{figure}

%\noindent
The experiment was performed with the ALADIN forward spectrometer
at the heavy-ion synchrotron SIS of the GSI Darmstadt, using a gold beam 
with an energy 
of 600~MeV per nucleon and a typical intensity of 2000 beam particles during 
a 500~ms spill.
A schematic view of the experimental setup in the bending plane of the ALADIN 
magnet is shown in figure \ref{setup}. 
The incoming beam entered the apparatus from the left and first hit the beam 
counters,
where for each beam particle the position in a plane perpendicular to the beam
direction and the arrival time were measured with resolutions 
$\delta_x \approx \delta_y \approx$ 0.5~mm FWHM and $\delta_t$
=100~ps FWHM, respectively. One meter downstream, the reaction target was 
positioned.
Targets of C, Al, Cu and Pb with 
a thickness between 200 and 700 mg/cm$^2$ were used, corresponding to
an interaction probability of up to 3\%. Light charged particles from the
mid-rapidity zone of the reaction  were detected by a Si-CsI-array which was 
placed at
angles between 7$^{o}$ and 40$^{o}$ with a solid angle coverage of 
approximately 30\%  in this angular range
(~50\% between 7$^{o}$ and 25$^{o}$, 15\% between 25$^{o}$ and 40$^{o}$~).
Fragments from the decay of the projectile spectator,
emitted into a cone of approximately 5$^{o}$ around the direction of the
incident beam,
entered the magnetic field of the magnet. The magnet was operated at a bending 
power of 1.4~Tm which corresponded to a deflection of 7.2$^{o}$ for fragments
with beam rigidity. The particles were 
detected in the TOF-wall, which was positioned 6~m behind the target. 
The time-of-flight of light particles with respect to the beam counter
was measured with a resolution of 300~ps FWHM and with a resolution 
of 140~ps FWHM for particles with a charge of 15 and above.
The TOF-wall provided the charge of all detected particles 
with single element resolution for charges up to  eight. 
Charged particles with charges of eight and above were simultaneously 
identified and tracked
by a time-projecting multiple-sampling ionization chamber TP-MUSIC II 
(~see section
\ref{SEC22}~), which was 
positioned outside the magnetic field
between the magnet and the TOF-wall. To minimize the influence of scattering,
of energy loss and of secondary nuclear reactions of the fragments after 
their production
in the target, the spectrometer up to an entrance window in front of the 
ionization chamber was 
operated in vacuum.
The components of the apparatus with the exception of the MUSIC detector have 
already been described in \cite{HUB91}. 

\subsection{The MUSIC detector}
\label{SEC22}

\begin{figure}
      \epsfxsize=8.0cm
      \centerline{\epsffile[50 100 490 470]{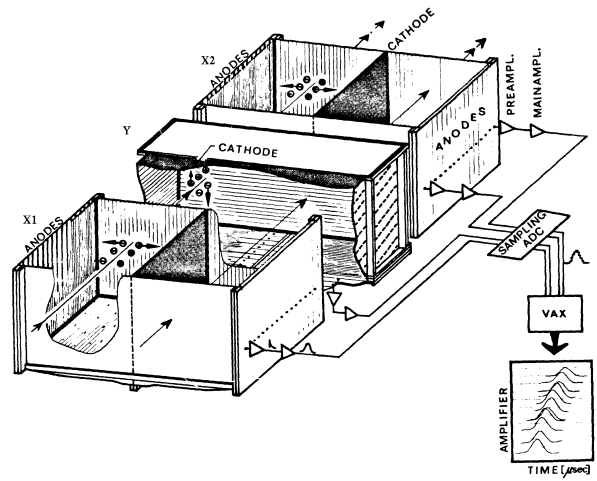}}
 \caption [] {\protect
  Schematic illustration of the design and the operation of the MUSIC II 
  detector. In the first and in the last cage, 
  the electric field is orientated horizontally, in the second field 
  cage vertically, providing the horizontal, the vertical and again the 
  horizontal track information. After amplification and pulse shaping the 
  signals are recorded by a sampling ADC. The insert shows 
  a raw time spectrum as delivered by the 16 anodes of each field cage
  for a track with a finite angle with respect to the anode plane.
 }
 \label{music}
 \vspace{0.1cm}
 \end{figure}

%\noindent
The TP-MUSIC detector is a time-projection multiple-sampling ionization 
chamber.
If a charged particle passes through its active volume, an ionization 
track containing positive ions -- which will drift to the 
cathode -- and free electrons -- which will move in the direction of the 
anodes -- is produced. 
Due to the homogeneous electric field, the drift velocity 
of the electrons towards the anodes is independent of the position within
the gas volume. Therefore,
the distance of the primary particle track from the anode is proportional to
the time the center of the electron cloud needs to reach the anode. 
The version TP-MUSIC~II \cite{BAU97} which was used in this experiment is 
shown in 
figure \ref{music} \footnote{The data published in \cite{SCH96,POC95,RUB96} 
were taken 
with the version III of the TP-MUSIC and a larger TOF-wall.}.
It consists of three active volumes with the drift field in adjacent sections
perpendicular to each other,
two for the measurement of the horizontal and one
for that of the vertical position and angle of the particle
track. Each field cage has an active area of 100~cm (horizontal) times 60~cm 
(vertical) and a length of 50~cm.
The horizontal field cages are both divided into two halves with a vertical 
cathode plane 
in the middle of the detector to
reduce the maximum drift length and the high voltages necessary to provide
the drift field. 
The chambers were operated at a high voltage of 150~V/cm, i.e.~7.5~kV for
the horizontal and 9~kV for the vertical field cages,  P10 (~90\% argon, 10\% 
methane~)
at a pressure of 800~mbar served as counting gas.
To allow multiple sampling of the particle signals, each
anode is subdivided into 16 stripes with a width of 3~cm
each. 

%\noindent
The anode signals were recorded using flash ADCs with a sampling rate of 
16~MHz. Together with a drift velocity of the electrons 
of approximately 5.3~cm/$\mu$s this corresponds
to amplitude measurements at a step size of 3~mm in the direction of the drift.
Since the drift time of the electron cloud  is measured
by each of the 16 anodes of a field cage  i.e.~at 16 points along 
the beam direction (z-direction),
the complete track information both in x- (~field cage 1 and 3~) and y-direction
(~field cage 2~) of the primary charged particle inside the 
MUSIC volume is available. The detector is operated outside the magnetic field 
volume 
of the ALADIN magnet, therefore the ionization track through the MUSIC gas is
a straight line which is obtained by fitting the 16 track positions by three
straight lines  - one in each field cage.

%\noindent
The position resolution has been estimated using 
the fact that the horizontal component of a track is determined with
two separate field cages. The intersections of the measured track segments 
from the 
first and the third field cage with a virtual reference plane, positioned
in the center of the vertical field cage and perpendicular to the z-direction,
are calculated. The distance between these two points of intersection is a
measure of the overall position and angle resolution of the detector.
Its distribution is a gaussian with a width of 2.4~mm FWHM for particles with 
a charge of 20 and above which increases to approximately 12~mm at the 
detection 
threshold of $Z=8$. These values are of the same order of magnitude as the
effect of small angle scattering of the fragments in the counting gas of 
the MUSIC.

%\noindent
The amplitude of the primary signal produced by a particular fragment is 
proportional to $ q^2\beta^2$, where $\beta $ is the velocity
of the particle and $q$ is its charge state. Fragments from the decay
of the projectile spectator are moving approximately with beam velocity.
In this case, all fragments with nuclear charges up to 50 are fully stripped 
after passing through the target matter. They remain fully stripped in the 
detector gas, the primary signal is therefore proportional to the square of 
the nuclear
charge of the particle. For particles with nuclear charges between 50 and 79, 
the mean charge exchange length in the MUSIC gas
(~8~cm and 30~cm for $Z=50$ and 79, respectively~) is small compared to the 
path length
of the particle within the MUSIC detector. They reach their equilibrium
charge state within the detector volume and the primary signal is 
proportional to the square of the effective charge. 

\begin{figure}
      \epsfxsize=8.0cm
      \centerline{\epsffile[0 100 560 420]{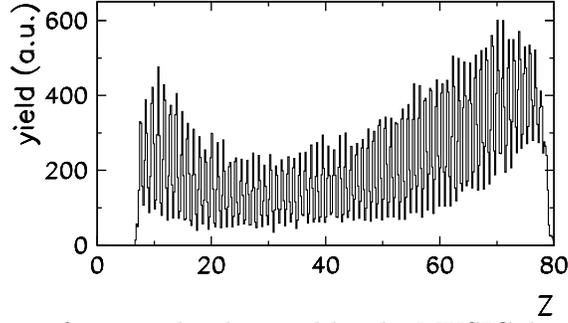}}
 \caption [] {\protect
  Charge spectrum for particles detected by the MUSIC detector in 
  the reaction Au+Cu. Single charge resolution is obtained for the whole
  range of fragments from the lower detection threshold Z=8 up 
  to beam particles. Fragments with charge 70 and above are suppressed
  by the trigger conditions.
 }
 \label{musicz}
 \vspace{0.1cm}
 \end{figure}

%\noindent
The amplitude of the primary
signal decreases due to diffusion broadening (~proportional to the drift 
distance~)
and due to impurities of the counting gas (~proportional to the square of the 
drift distance~). 
The amplitude measured at the anode is therefore
dependent on the drift distance of the electron cloud. 
To determine the position correction the incident beam --~i.e.~particles with 
known $Z$ and $\beta $~-- is swept across the field cages by varying the 
field of 
the magnet. In addition,
the signals are corrected for the deviations
from the beam velocity. This is essential for the charge resolution of binary
fission fragments which have the widest distribution of laboratory velocities
of all heavy fragments ($Z \geq 8$) from the decay of the projectile spectator.
A charge resolution of 0.5 charge units FWHM is reached. This
is demonstrated in figure  \ref{musicz} where a charge spectrum 
of the MUSIC-detector is shown. Since both the differences in pulse height for 
two neighboring charges and their fluctuations are proportional 
to $Z$, the charge resolution is independent of the charge of the fragment.
The lower threshold for particle identification reached in this experiment 
is $Z=8$.

\subsection{Momentum and mass reconstruction}
\label{SEC23}

%\noindent
From the tracks of the charged particles measured behind the ALADIN magnet, 
the rigidity 
vector can be determined if the magnetic field is known. It was decided to 
fit the 
particle properties as a function of the measured track parameters rather 
than using 
a backtracing method because the latter is more time consuming at the 
analysis stage. 
For particles with a rigidity vector $\vec{R}$, trajectories starting at the 
target position $z_{targ}$ and coordinates ($x_{targ}$, $y_{targ}$) within 
the beam spot
are calculated using the routines provided by the program package GEANT 
\cite{GEA}. 
The starting conditions are chosen from a five-dimensional grid with 
equidistant spacing
for the variables $x_{targ}$, $y_{targ}$, $1/R$, $R_{x}/R$ 
and $R_{y}/R$. For a given magnetic field strength of the ALADIN magnet,
the intersection ($x_{music}$, $y_{music}$) of each track with the reference
plane of the MUSIC detector and its angle ($m_x$, $m_y$) relative to this plane 
as well as the path length to this point are 
determined. The bending plane of the magnet is 
the horizontal x-z-plane, i.e.~the main component of the
magnetic field points to the direction of the y-axis, although the fringe 
fields can not
be neglected, especially if the full geometric acceptance is used.

%\noindent
Since a large range in $N/Z$-ratios (0.7 - 1.5) and emission angles has to be
covered, only 40\% of the grid points correspond to trajectories which reach 
the 
reference plane behind the magnet, all others end at the wall of the magnet
chamber where they are lost. For the successful tracks, the three components 
of the rigidity 
vector together with the path length are fitted as the product of 
one-dimensional 
functions of five variables:
the position ($x_{music}$, $y_{music}$), the angle $m_x$
and the target position ($x_{targ}$, $y_{targ}$). The fit is done by means of an
expansion in series of Chebychev polynomials for each variable. 
For a magnet which
has virtually a dipole field, the most relevant terms are linear in 
$x_{music}$ and $m_x$
for $1/R$, $R_{x}/R$ and the path length, and linear in $y_{music}$
for $R_{y}/R$, but for a accuracy of the momentum reconstruction on the percent
level, higher order terms can not be ignored. Under the assumption of 
an expansion up to
third order, approximately 1000 individual contributions have to be calculated,
which is not feasible. However, a particular term can be estimated by 
the size of 
the related expansion coefficient, since Chebychev polynomials 
are orthogonal within the interval from -1 to 1, and at the same time all their 
minima and maxima within this interval have the values -1 and 1, respectively.
(~Strictly mathematically speaking, this is not correct. Among other 
conditions, the orthogonality
relations can only be used if the full parameter space is covered. This 
is not the case,
since not all of the tracks reach the reference plane.~)
In a second step, small terms are gradually
suppressed until the $\chi ^2$ of the fit has increased by 10\%, 
thus reducing the total number from between 400 to 1000 in the first step
(~depending on the highest order taken into account~) to 25 to
40 (~depending on the variable~). The fitting procedure is then repeated 
using only 
the remaining relevant terms which leads to slightly
different expansion coefficients in the final result.

%\noindent
Once this fitting procedure has been
performed for each setting of the magnetic field used during the experiment, 
the reconstruction of the rigidity vector 
and of the path length is reduced to the evaluation of a set of polynomials.
If a reasonable qualitiy of the reconstruction can be archieved, this is a 
justification for the somewhat heuristic method to select the relevant 
contributions.
The accuracy can easily be determined by calculating tracks
with random start values --~i.e.~with 
starting coordinates at the target and for the rigidities 
not identical with the 
starting parameters used for the fitting procedure.
The reconstruction is done for these tracks by evaluating the fit functions, 
and the input values are compared 
to the reconstructed ones. The mean deviations for rigidity and
path length are a measure of the uncertainty caused by the reconstruction 
method itself.
Clearly, the size of these deviations is
on the one hand dependent on the mesh size of the grid of start values and on 
the other
hand on the choice of the highest order taken into account for the expansion in 
Chebychev polynomials.
Both quantities were optimized 
until the internal accuracy for all variables was better than 0.1\% FWHM 
within the
chosen range of rigidities between 1.2 and 3.6 GeV/c. The final set of 
coefficients was 
obtained by fitting $\approx$12000
tracks with a maximum order of 4 for each polynomial and a maximum of 6 
for the sum of the orders within a term.

\begin{figure}
      \epsfxsize=8.0cm
      \centerline{\epsffile[0 0 300 300]{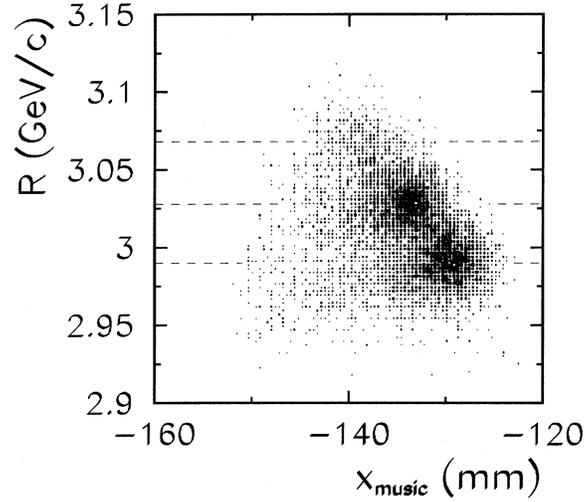}}
 \caption [] {\protect
  Rigidity $R$ of gold projectiles at 600~MeV/nucleon 
  versus the horizontal position
  in the reference plane of the MUSIC detector. Only events were
  selected where the gold nuclei passed through the carbon target
  without any nuclear interaction. The three charge states $79^{+}$,
  $78^{+}$ and $77^{+}$ correspond to the rigidities 2.990, 3.028 
  and 3.068 GeV/c, respectively, which are different by 1.3\% (dashed lines).
 }
 \label{beam}
 \vspace{0.1cm}
 \end{figure}

%\noindent
A very similar procedure as described above 
can be used to estimate the expected errors due to the experimental resolution 
of the two position detectors in front and behind the magnet. A random 
offset of the order of the experimental 
uncertainties is added  to the positions and slopes prior to the
evaluation of the polynomials. Afterwards,
the difference between the reconstructed values with and without 
random offsets is calculated. The mean value of these deviations
is the resolution expected due to the experimental uncertainties. It was 
found that
both an uncertainty of 0.7~mrad 
and of 3~mm produce an error in the rigidity of 1\%. With the time and position
resolutions given in the previous section, rigidity resolutions of 
approximately 
1.2\% and 3\% can be expected for beam particles and medium heavy 
fragments with charge $\approx $12, respectively. 
The quality of the rigidity reconstruction can be demonstrated by the 
rigidity distribution of
beam particles passing through the apparatus without any nuclear interaction.
Within all targets used in this experiment, gold projectiles reach their 
equilibrium charge
state, providing particles with identical momenta and charge states 77$^+$, 
78$^+$, 79$^+$,
i.e.~with rigidities which differ by 1.3\% per charge state.
With the carbon target, the influence of angular 
straggling within the target is small and negligible compared to the 
experimental errors due to the position resolution.
In figure \ref{beam}, the rigidity distribution of beam particles after 
passing through 
the carbon target is plotted versus their x-position in the MUSIC reference 
plane. 
In this representation of the data,
the three charge states (~ equilibrium charge state distribution of 600~MeV/u 
gold in carbon:
59\% of the projectiles are fully stripped, 35\% have a charge state of 78$^+$ 
and
6\% of 77$^+$~ \cite{STO91}~) are clearly visible, i.e.~the rigidity resolution 
for heavy nuclei is approximately 1.3\% FWHM
which is in agreement with values expected from the resolutions of the 
individual detectors.

%\noindent
Using the reconstructed values for the rigidity and path length, 
the charge of the particle measured by the MUSIC detector and 
the time of flight given by the TOF-wall,
the velocity and the momentum vector can be calculated for each charged 
particle detected
both in the MUSIC and the TOF-wall, i.e.~for particles
with a charge of eight and above. 
The knowledge of velocity and momentum allows the calculation of the 
particle's mass.

\begin{figure}
      \epsfxsize=8.0cm
      \centerline{\epsffile[0 60 490 540]{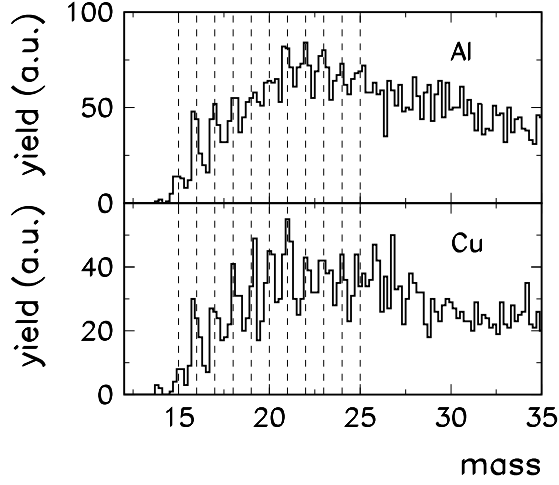}}
 \caption [] {\protect
  Mass spectrum of particles detected in the MUSIC detector for the
  systems Au+Al and Au+Cu. Single mass resolution is obtained for charges
  up to 12.
 }
 \label{masses}
 \vspace{0.1cm}
 \end{figure}

%\noindent
In figure \ref{masses}, the mass spectra 
for the reactions Au+Al and  Au+Cu are shown. 
Single mass resolution for charges up to 12 is obtained, corresponding to
a mass resolution $\Delta$A/A of approximately 4.0\% FWHM for light fragments.
The dominant contribution to the uncertainty of the mass measurement
\begin{equation}
  \frac{\Delta A}{A} = 
  \sqrt{ \left( \frac{ \Delta R}{R} \right) ^{2} + \left( \gamma^{2} \cdot 
  \frac{ \Delta TOF}{TOF} \right) ^{2} } \quad .
\end{equation}
is caused by the mass dependent error of the time measurement which is 
amplified by the factor
$\gamma^2$ (~$\gamma^2$=2.6 for 600~MeV/nucleon~). From this, a rigidity 
resolution of 2.4\%
FWHM can be deduced for light fragments.

\section{Data}
\label{SEC3}

%\noindent
The breakup dynamics of multifragmenting
spectator matter will be reflected in the momenta of the fragments produced.
Especially observables combining the kinematic information of two or more 
particles
--~e.g.~relative velocities~-- are governed by the long-range Coulomb 
repulsion and are therefore 
sensitive to time scales of the decay and spatial properties of the 
decaying source. Clearly, the 
breakup pattern will change with increasing 
excitation energy transferred to the spectator matter. 
From the analysis of reactions of gold projectiles with different 
targets and beam energies
between 400 and 1000~MeV/nucleon it is well established 
\cite{HUB91,SCH96,POC95} that
the quantity $Z_{bound}$ --~defined as the sum of the charges
of all particles with charge two and above, which are emitted from the 
projectile spectator
and detected in the TOF-wall~-- reflects directly the size of the 
spectator as well as 
the excitation energy transferred to the spectator 
nucleus. It was furthermore
shown  that the mean number of fragments produced in a reaction as well 
as other observables
characterizing the populated partition space were independent of the 
target used, if they were
investigated as a function of the quantity $Z_{bound}$.
$Z_{bound}$ is therefore used as a sorting parameter describing the 
violence of the reaction.

%\noindent
As was discussed earlier, momenta and masses could only be reconstructed 
for particles with
a charge $Z$ of eight and above. It will be shown in the next section that 
events with two and more
large fragments with $Z \geq 8$ cover the $Z_{bound}$ range from 30 to 70. 
The maximum mean number of 
intermediate mass fragments --~defined as fragments with charges between 3 
and 30~-- is observed for
a $Z_{bound}$ value of approximately 40. The dataset available covers therefore 
the range from peripheral collisions up to the region of maximum fragment 
production.

\subsection{Characterization of two- and three-particle events}
\label{SEC31}

\begin{figure}
      \epsfxsize=8.0cm
      \centerline{\epsffile[0 30 540 540]{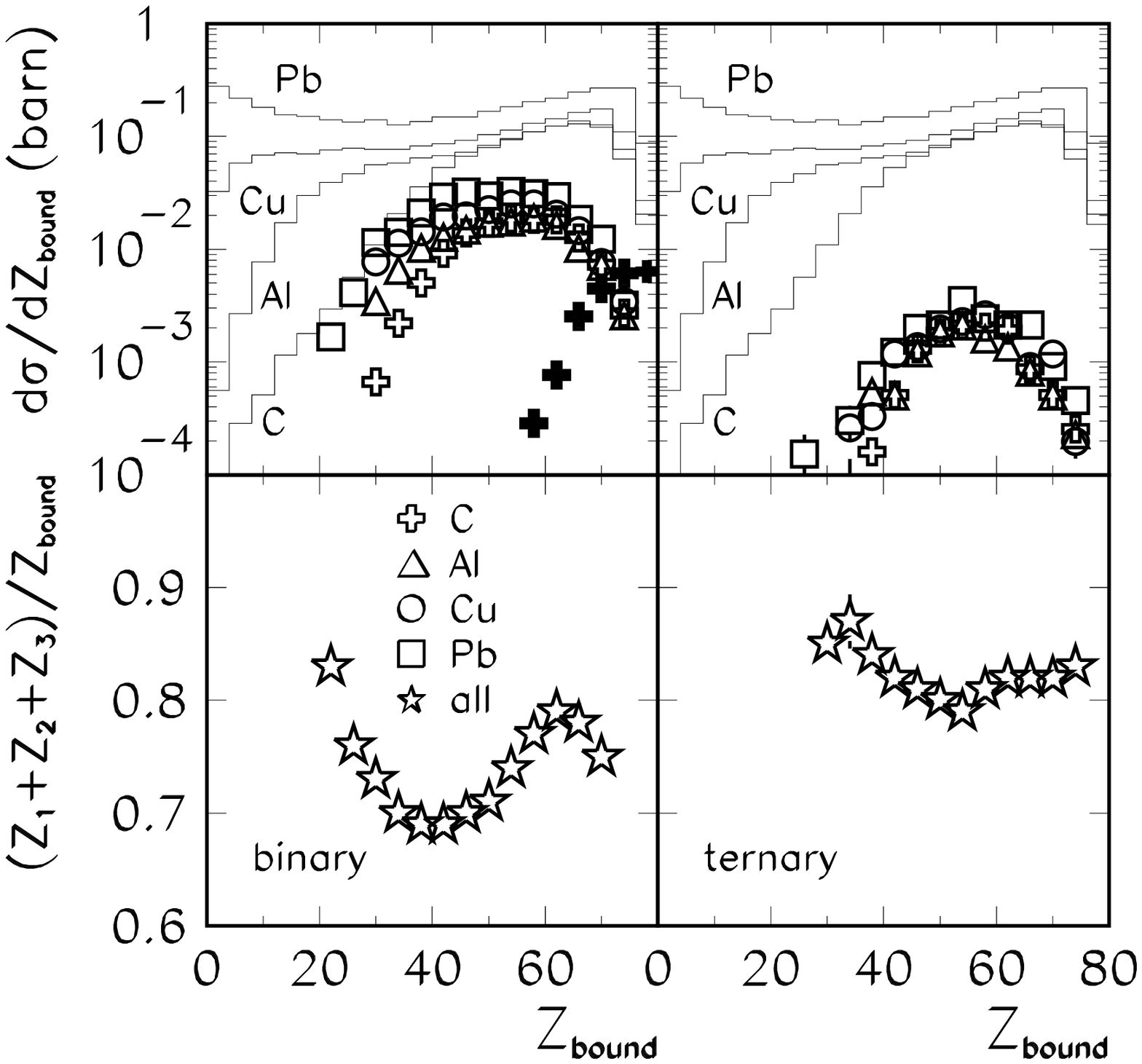}}
 \caption [] {\protect
  (top): Differential cross sections d$\sigma$/d$Z_{bound}$ both for the 
  inclusive 
  data (histograms) and for events with two and three fragments of charge 
  $\geq$8 in the exit channel. 
  For the system Au+C the cross section for binary fission is also shown (solid 
  crosses). \\
  (bottom): Fraction of $Z_{bound}$ contained in the sum of the charges of the 
  two or three heavy fragments versus $Z_{bound}$. The symbols represent a cross
  section weighted mean value for all four targets.
 }
 \label{wqternary}
 \vspace{0.1cm}
 \end{figure}

%\noindent
To show the characteristics of the event classes with two and three heavy 
particles with
charge $Z \geq 8$, their reaction cross sections $d\sigma /d Z_{bound}$ 
are plotted in the upper panel of figure \ref{wqternary} for the four 
different targets
as a function of $Z_{bound}$. In the following, events with two (three) 
fragments with 
charge $Z \geq 8$ are called binary (ternary).
Binary events attributed to binary fission were excluded by the condition 
that either
the lighter fragment is of charge below 20 or the sum of the two charges is
smaller than 60. In the $Z_1 Z_2$-plane, this region is well separated
from the region of binary fission \cite{RUB96}.
For comparison, the inclusive reaction cross sections --~i.e.~without 
conditions on 
fragment multiplicity and charge~-- are also shown: The binary and 
ternary events 
represent approximately 10\% and 1\%  of the nuclear reaction cross 
section, respectively.
In order to demonstrate that binary fission events as defined above
populate an impact parameter region different from that of binary events 
without fission,
the cross section for binary fission in the reaction Au+C is included. 
These events 
are obviously produced in very peripheral reactions. It had been shown 
earlier \cite{KRE93}
that multifragment events evolve --~with decreasing $Z_{bound}$~-- from 
events with one heavy 
residue in the exit channel of the reaction and not from binary fission events.

%\noindent
The sum of the charges of the two and three fragments is plotted in the lower 
panels. In ternary events, typically 80\% of $Z_{bound}$ is contained in 
the charges of the three
heavy fragments with average charges and masses of 
$\langle Z_i\rangle$ = 22, 13, 10 and
$\langle A_i\rangle$ = 48, 29, 20, i=1,2,3. In binary events, the sum of the
charges of the heavy fragments accounts on average for 75\% of Z$_{bound}$ with 
a clear minimum at Z$_{bound}$=40,
where the maximum mean number of intermediate mass fragments is observed. 
The average charges and masses for this event class are
$\langle Z_i\rangle$ = 26, 13 and $\langle A_i\rangle$ = 57, 27, i=1,2.

%\noindent
It will now be demonstrated that these two event classes are representative 
subsets of 
the experimental data, i.e.~that for a given $Z_{bound}$ value no evidence 
for a 
strong dependence on the 
number of heavy fragments is found. This means that 
other quantities defining an event do not show a close correlation between 
their mean values
and  the multiplicity of the heavy fragments if analyzed according to 
$Z_{bound}$.
Evidently, only observables can be used for this investigation which are 
not dominated 
by autocorrelations. The multiplicity of intermediate mass 
fragments for instance contains the number of all 
heavy fragments with a charge smaller than 30.
The mean multiplicity of IMFs is therefore influenced by the selection 
criterion and
will be significantly different for events
with different numbers of heavy fragments in the exit channel.

\begin{figure}
      \epsfxsize=8.0cm
      \centerline{\epsffile[20 10 490 540]{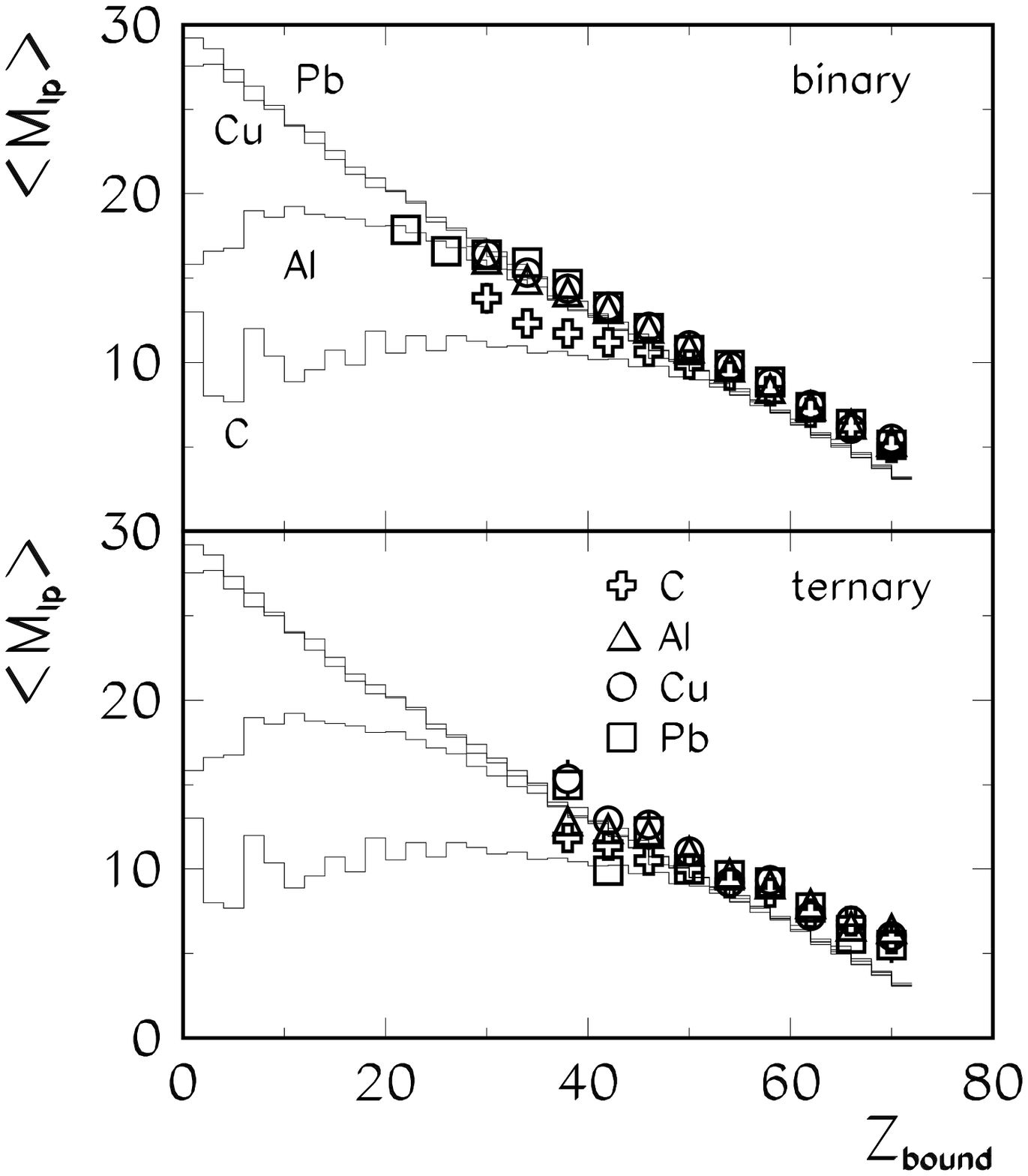}}
 \caption [] {\protect
  Mean multiplicity $\langle M_{lp} \rangle$ 
  of light particles from the interaction zone of the reaction 
  detected in the hodoscope versus 
  $Z_{bound}$. The histograms show the inclusive distributions for the 
  four targets. \\
  (top): Binary events where binary fission -- as defined in the 
  text -- was excluded. \\
  (bottom): Ternary events \\
  $\langle M_{lp} \rangle$ decreases with increasing size of the projectile 
  spectator.
  Within the experimental errors, the multiplicity for a given value of 
  $Z_{bound}$ is independent of the number  of the projectile 
  fragments. This holds for the whole range of $Z_{bound}$  with the exception 
  of very peripheral reactions with 
  $Z_{bound} \leq$65 where the inclusive distributions are dominated by 
  spallation reactions.
 }
 \label{multlp}
 \vspace{0.1cm}
 \end{figure}

%\noindent
The mean number $\langle M_{lp}\rangle$ of light particles from the 
mid-rapididy
zone of the reaction which were detected in the hodoscope 
is a quantity which is certainly dependent on the violence of the reaction but
independent of the specific decay channels of the excited projectile spectator.
In figure \ref{multlp}, the inclusive distributions of $\langle M_{lp}\rangle$
versus $Z_{bound}$ for the four targets are shown together with the 
distributions
for events with two and three heavy particles. 
In agreement with the participant-spectator-model, the size of the 
interaction zone --~represented by the mean number of light particles~--
increases with decreasing size of the projectile spectator. The distributions
are independent of the multiplicity of heavy projectile fragments with 
the exception of 
the most peripheral reactions (~$Z_{bound} \geq$ 65~). 
In this range of largest impact parameters the inclusive data are dominated by
spallation and not by multifragmentation events. There, the restriction to
events with two or three heavy particles in the exit channel is 
synonymous with the selection of events with higher mean energy.

\begin{figure}
      \epsfxsize=8.0cm
      \centerline{\epsffile[10 50 480 540]{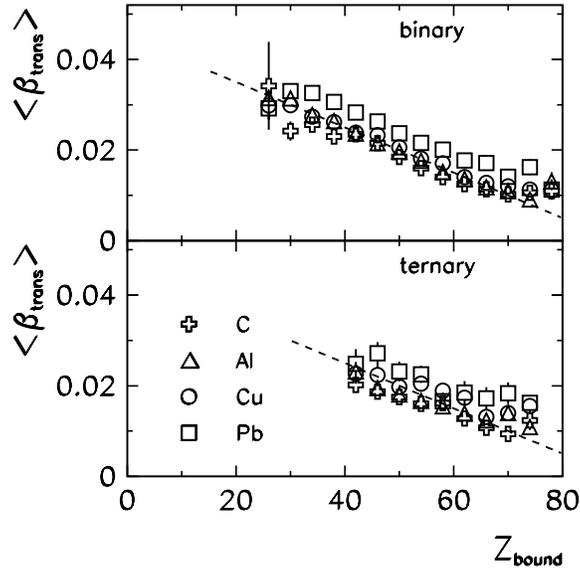}}
 \caption [] {\protect
  Mean transversal velocity $\beta _{trans} = \sqrt{\beta_x^2 + \beta_y^2}$ of 
  the center of mass 
  relative to the velocity of the beam for events 
  with two (upper panel) and three (lower panel) heavy particles with 
  $Z_{i}\geq$8 in the exit channel. The bounce decreases with decreasing
  centrality of the reaction. The non-zero value for the largest $Z_{bound}$
  bins is due to the trigger condition which requests the detection of at least 
  one light particle in the hodosope. The dashed lines in both panels show 
  the same linear fit to the binary data, to demonstrate that 
  for a given value of $Z_{bound}$ the distributions do not depend on the
  multiplicity of heavy particles.
 }
 \label{betaparperp}
 \vspace{0.1cm}
 \end{figure}

%\noindent
The transversal deflection of the decaying projectile spectator is 
another quantity which 
is not influenced by autocorrelations with regard to the decay pattern.
Since in events with two or three heavy particles in the exit channel the heavy
particles contain typically 75-80\% of $Z_{bound}$, the center of mass of these
particles is in good approximation the center of mass of the decaying system.
Thus, the transversal velocity 
\begin{equation}
  \beta _{trans} = \sqrt{\beta _{x}^{2} + \beta _{y}^{2}}
\end{equation}
of the
center of mass of the two or three particles with respect to the beam frame 
was calculated. In figure \ref{betaparperp}, the mean values of this 
velocity as a function of $Z_{bound}$ are compared for 
events with two and three heavy fragments in the exit channel.
In agreement with inclusive measurements at 400~MeV/nucleon \cite{KUN94},
$ \beta _{trans}$ increases monotonously with decreasing $Z_{bound}$ 
and establishes
the transversal deflection of the projectile spectator and therefore the
transversal momentum transfer (bounce) 
as a measure of the deposition of 
excitation energy into the spectator matter.
Pure Coulomb interaction during a grazing collision would lead to very 
small values for
the bounce between $5 \cdot 10^{-4}$c and $ 4 \cdot 10^{-3}$c for C and Pb, 
respectively.
But due to the trigger condition 
demanding at least one light particle detected in the hodoscope and therefore
a nuclear reaction, the bounce does not vanish for $Z_{bound} = 80$.
The increasing Coulomb repulsion with increasing charge of the target nucleus
is nevertheless reflected in the small target dependence.
Within the experimental errors, the transversal velocity at a given $Z_{bound}$
is independent of the two decay patterns studied.

%\noindent
The two quantities $\langle M_{lp}\rangle$
and $ \beta _{trans}$  describe properties related to the initial reaction phase
--~the size of the fireball and the excitation energy transferred to the 
spectator matter. The fact that these quantities are independent of a 
specific choice 
of the multiplicity of heavy fragments demonstrates that a restriction to 
the subset 
of events, defined by the detection threshold of the MUSIC detector,
does not select a non-typical sample of the produced projectile spectators.

\subsection{Two- and three-particle observables}
\label{SEC32}

%\noindent
From the measured momenta of the heavy fragments the intrinsic 
momenta $\vec{p}_{cm}(i)$ and velocities $\vec{v}_{cm}(i)$ 
in the center of mass frame (CM-frame) of the binary or ternary heavy 
fragment system werde determined. This has 
primarily the advantage of eliminating the projectile velocity from the 
analysis. 
Furthermore, it reduces the influence of directed collective motion on the
momenta of the particles. This is especially
important if the data are to be compared to calculations with models which do
not include linear collective motions. By construction, these momenta are
collinear
in the case of two and coplanar in the case of three particle events. For the
further
analysis, a new coordinate system has been chosen such that for each event 
the momentum vectors lie in the same plane -- the xy-plane -- 
and that the direction of the heaviest particle coincides with the 
x-axis. This eliminates the three Euler angles which describe the spacial 
orientation
of the momenta relative to the beam axis.
The kinematics of the two and three heavy fragments is thus 
reduced to one (~$p_{x}(1)$~) and three (~$p_{x}(1)$, $p_{y}(2)$, 
$p_{x}(2)-p_{x}(3)$~) parameters, respectively. The 
relative kinematics of the fragments can thus uniquely be expressed in terms 
of one and three independent quantities which, for the 
analysis presented in this paper, are chosen as follows:
(i)~the total kinetic energy $E_3$ of the fragments in the CM-frame,
(ii)~the reduced relative velocity $v_{red}(2,3)$, 
and (iii)~a quantity $\Omega_{\Delta}$ 
which describes the event shape
in velocity space. In the case of only two heavy particles, the kinetic energy
$E_2$ alone is sufficient to describe the decay dynamics.

\begin{figure}
      \epsfxsize=8.0cm
      \centerline{\epsffile[20 40 560 530]{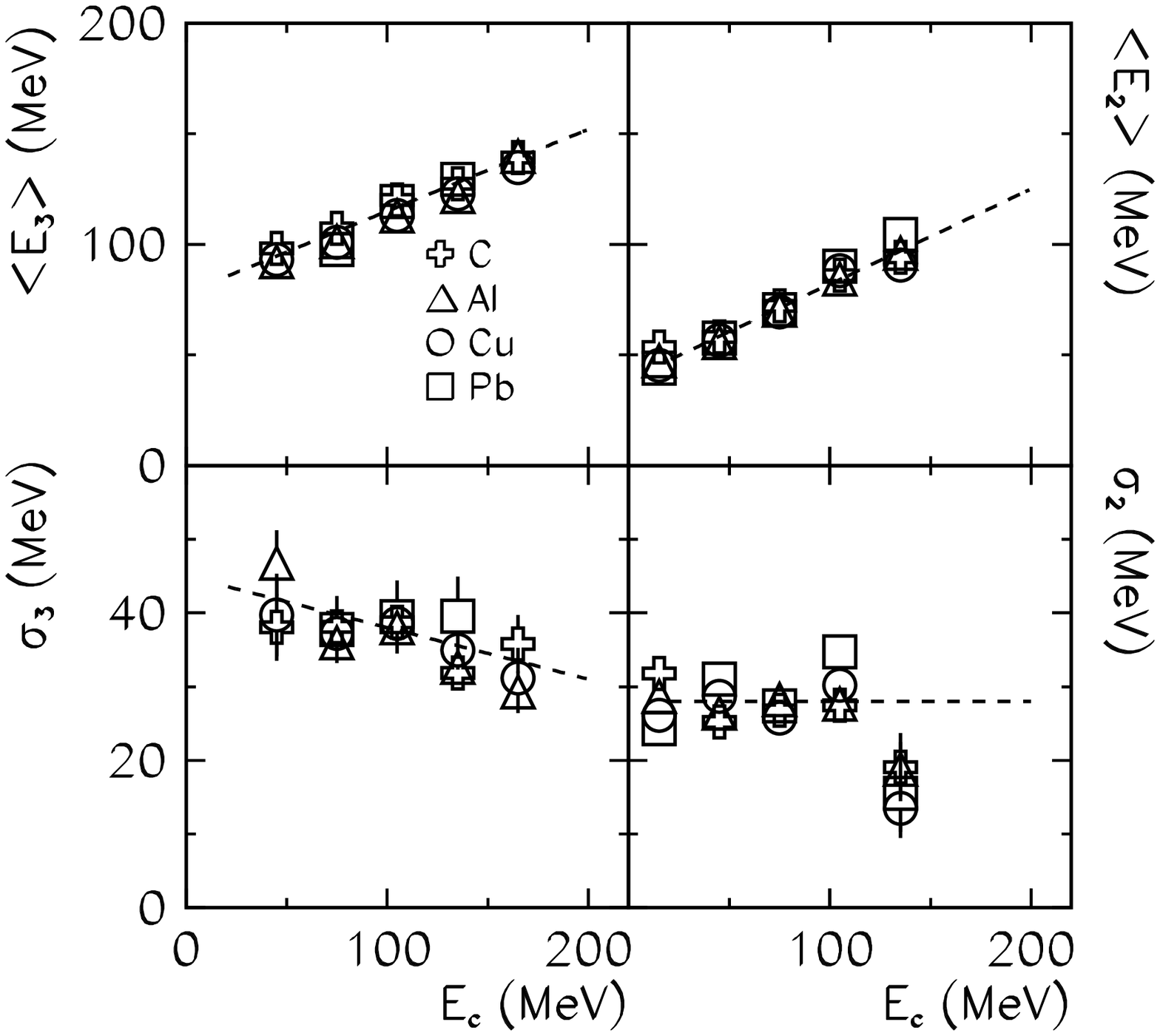}}
 \caption [] {\protect
  (left panels): Mean value 
  $\langle E_3 \rangle$ (top) and the standard deviation 
  $\sigma_3$ (bottom) of the total kinetic energy 
  as a function of the nominal Coulomb energy $E_c$ for events with three
  large fragments in the exit channel of the reaction. \\
  (right panels):  $\langle E_2 \rangle$  (top) and the standard deviation
  $\sigma_2$ (bottom), the equivalent quantities for events with two large
  fragments where binary fission is excluded. \\
  The results are shown for the systems  Au+C, 
  Al, Cu and Pb at E/A = 600 MeV.
  Both the mean kinetic energy and the width of the energy distribution 
  are within the experimental errors independent of the target.
  The straight lines are least square fits to the combined data of all targets. 
  }
 \label{ekin3}
 \vspace{0.1cm}
 \end{figure}

%\noindent
The sum $E_3$ of the kinetic energies of the three particles is calculated 
in their 
CM-frame 
\begin{equation}
  E_3 = \sum_{i=1}^3 \frac{p^2_{cm}(i)}{2\cdot m_0 \cdot A_i}  \quad ,
  \label{EQ_2}
\end{equation}
where $m_0$=931.5 MeV/c$^2$ is the atomic mass unit and $A_i$ the mass number
of the fragment $i$. The kinetic energy of the particles is dominated by the 
Coulomb
interaction which itself is strongly dependent on the charges involved.
The mean value $\langle E_3 \rangle $ is therefore studied together with 
the standard deviation $\sigma_3$ of the $E_3$-distribution
as a function of the nominal Coulomb-repulsion $E_c$ of the fragments at the 
time
of the breakup, i.e.~as a function of the Coulomb potential 
of three touching spheres
with radii $R_i$ = 1.4 $\cdot$ $A_i^{1/3}$: 
\begin{equation}
  E_{c} = e^2 \cdot \sum_{i<j} \frac{Z_i \cdot Z_j} 
          {1.4\cdot (A_i^{1/3}+A_j^{1/3})} \quad .
  \label{EQ_3}
\end{equation}
This is a generalization of the well known Viola formula \cite{VIO63}.
For events with only two heavy fragments the kinetic energy and the Coulomb 
repulsion 
are calculated accordingly. 
The experimental results 
are plotted in figure \ref{ekin3} for the four targets used. 
Within the statistical uncertainties, no target dependence is apparent.
In all further plots, mean 
values of the kinetic energy and of the width of the energy distribution for the
combined data of all four targets will therefore be shown. 
$\langle E_2 \rangle$, $\langle E_3 \rangle $, and $\sigma_2$, $\sigma_3$ 
depend linearly on $E_{c}$ and are parameterized in terms of
straight line fits ($y=m \cdot x+b$) common to the data of all four targets. 
The slopes and intercepts of these fits are listed in the following table:
\begin{center}
\begin{tabular}{|l||rl|rl|} 
     \hline
           & \multicolumn{2}{c|}{binary} & \multicolumn{2}{c|}{ternary} \\
     \hline \hline
   m$_E$               &  0.43 & $\pm$ 0.05 &   0.37  & $\pm$ 0.04 \\
   b$_E$ (MeV)         & 39.0  & $\pm$ 4.0  &  76.0   & $\pm$ 5.0  \\
   m$_{\sigma}$        &  0.0  & $\pm$ 0.05 &  -0.07  & $\pm$ 0.01 \\
   b$_{\sigma}$ (MeV)  & 28.0  & $\pm$ 3.0  &  44.0   & $\pm$ 4.0  \\
     \hline
\end{tabular}
\end{center}

%\noindent
The parameters $b_E$ and $b_{\sigma}$ 
describe the mean energies and their variations  in the limit of $E_{c}$=0,
i.e.~without Coulomb interaction, both for binary
and ternary events. Under the assumption of a purely thermal source with a 
temperature 
$T$ and without Coulomb interaction
the mean values $\langle E_2 \rangle $ and $\langle E_3 \rangle $ of the
kinetic energy distributions are $2T$ with a width $\sigma_2$ of
$\sqrt{2} T$ and
$4T$ with a width $\sigma_3$ of $2T$ in case of surface emission and $3/2 
\cdot T$ with 
a width $\sigma_2$ of $\sqrt{3/2} T$ and $3T$ with a $\sigma_3$ of $\sqrt{3} 
T$ in
case of volume emission of the fragments. For both breakup scenarios, the 
temperatures
deduced from these relations are within the experimental errors identical for
binary and ternary events. The assumption of volume emission leads to a 
temperature
of 25~MeV whereas the value for surface emission is 20 MeV.
Results obtained in the reaction Au + Au at 1000~MeV/nucleon
where kinetic temperatures were extracted from the energy spectra of light 
charged
particles up to $^4$He emitted from the target spectator \cite{XI97} and
temperatures extracted from transverse momentum distributions at 600~MeV/u 
\cite{SCH96}
are of similar size (~15-20 MeV~).

%\noindent
In line with previous studies \cite{KIM92}, the reduced relative 
velocity 
is defined as
\begin{equation}
  v_{red}(i,j)=\frac{v_{rel}(i,j)}{\sqrt{Z_i+Z_j}} \quad ,
  \label{EQ_1}
\end{equation}
where $v_{rel}(i,j)$ is the relative velocity of particles $i$ and $j$ 
and $Z_i$ and $Z_j$ are the corresponding charges of the fragments. 
With this definition, the mutual Coulomb repulsion within a fragment pair is 
charge independent. For ternary events, the reduced relative 
velocity of the second and third largest fragment is calculated.
Its mean experimental value, averaged over all targets,  is 
0.0206$\cdot$c $\pm$ 
0.0005$\cdot$c. This value will be used later on to adjust the input parameters
of model calculations.

\begin{figure}
      \epsfxsize=8.0cm
      \centerline{\epsffile[0 0 550 510]{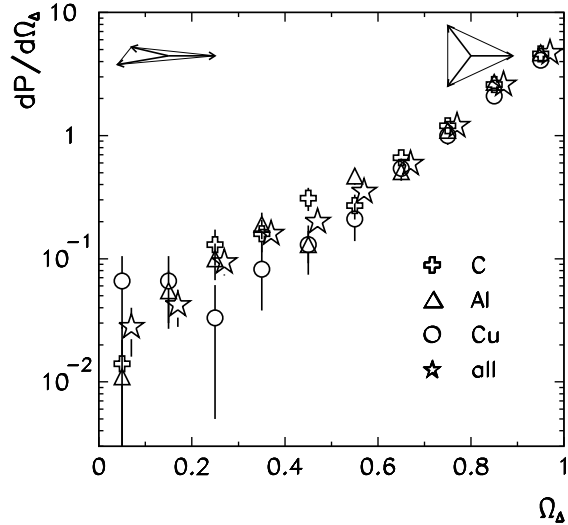}}
 \caption [] {\protect
  Experimental probability distributions for $\Omega_{\Delta}$. The 
  symbols for the combined data of all targets are shifted by 0.02 in 
  x-direction. The event shape in velocity space for large and small
  values of $\Omega_{\Delta}$ are indicated schematically in the
  upper left and right corners.
 }
 \label{omega}
 \vspace{0.1cm}
 \end{figure}

\begin{figure}
      \epsfxsize=8.0cm
      \centerline{\epsffile[0 0 540 510]{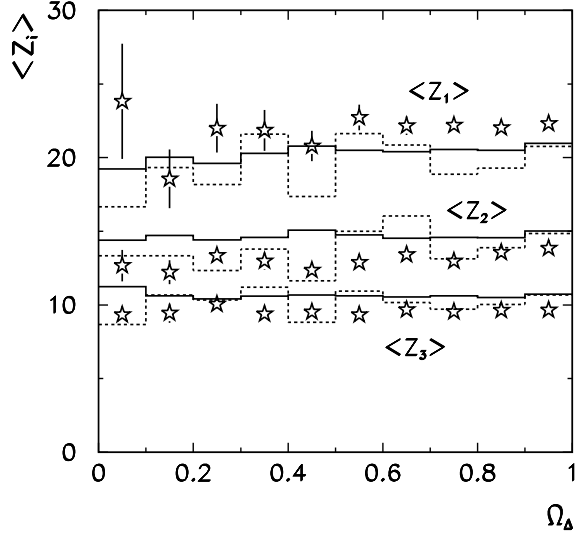}}
 \caption [] {\protect
  Mean values of $Z_i$ (i=1,2,3) as a function of $\Omega_{\Delta}$ for the 
combined
  data of all targets (symbols). 
  Also included are the results of calculations with the SOS-model assuming a
  sequential (solid histograms) and a simultaneous (dashed histograms) decay
  of the projectile spectator (~see section \ref{SEC33}~).
  For all three fragments, the mean charges are independent of 
  $\Omega_{\Delta}$,
  showing that the probability distribution for $\Omega_{\Delta}$ 
  is not a trivial 
  consequence of the charge distributions.
 }
 \label{z123}
 \vspace{0.1cm}
 \end{figure}

%\noindent
The third quantity $\Omega_{\Delta}$ characterizes the configuration of the 
three velocity vectors
\begin{equation}
  \Omega_{\Delta} = \frac{ \Delta_{123} }{ \Delta_0}  \quad ,
  \label{EQ_4}
\end{equation}
where $\Delta_{123}$ denotes the area of the triangle with its three sides 
given by the
three relative velocities $\vec{v}_{rel}(1,2)$,  $\vec{v}_{rel}(2,3)$, and 
$\vec{v}_{rel}(1,3)$.
The normalization $\Delta_0$ represents the area of an equilateral triangle 
with an circumference of 
\begin{equation}
  u = |\vec{v}_{rel}(1,2)| + |\vec{v}_{rel}(2,3)| + |\vec{v}_{rel}(1,3)| \quad ,
\end{equation}
which is the largest area possible for a given circumference.
Thus, $\Omega_{\Delta}$ varies between 0 and 1, where
$\Omega_{\Delta}$=0 corresponds to a streched configuration with the three
relative velocities being collinear and 
$\Omega_{\Delta}$=1 corresponding to a situation where the three
CM-velocities point to the corners of an equilateral triangle.
The normalized experimental distributions of the reduced area $\Omega_{\Delta}$
are shown in figure \ref{omega} for the four targets:
The probability to find an equilateral velocity configuration
is two orders of magnitude larger than that for a stretched one.
Within the statistical errors, the distributions are independent of the target, 
therefore the mean value averaged over all four targets was determined to 
increase the statistics especially for small values of $\Omega_{\Delta}$.
In order to address the question of possible 
correlations between the event shape and the charges of the
fragments, the average charges $\langle Z_i \rangle $ ($i$=1,2,3) 
of the three fragments ordered according to their sizes
are studied as a function of $\Omega_{\Delta}$
for the combined data of all targets.
The results are shown in figure \ref{z123}. 
Within the statistics, the average charges are independent of 
$\Omega_{\Delta}$, indicating that the probability
distribution of $\Omega_{\Delta}$ is not
a trivial consequence of the charge distribution or the spectator size.

\subsection{Sensitivity of the three-particle variables}
\label{SEC33}

%\noindent
In order to illustrate the potential sensitivity of the chosen observables,
calculations with the schematic SOS-code \cite{LOP92} were performed.
This code was especially developped to study 
the influence of two extreme
breakup mechanisms on experimentally observable kinetic quantities,
using in both cases a nuclear system of a 
given size and excitation energy and identical multifragment channels.
It produces multifragment events 
with two sets of momentum distributions, simulating for each event
on the one hand a sequence of binary decays
and on the other hand a simultaneous breakup using the final partition 
of the sequential decay chain and placing the fragments randomly but without
overlap in a sphere.

%\noindent
For this investigation, masses and excitation energies of the 
decaying spectator nuclei were chosen according to \cite{BAO93} where  the 
authors
adjusted the input parameters of a statistical fragmentation model (Berlin 
model) until the
relation between $\langle M_{imf}\rangle$ and $Z_{bound}$ was well reproduced 
for
the system Au + Cu at 600~MeV/nucleon.
Since the main motivation of the calculations using the SOS code was 
to illustrate the potential usefulness of the presented 
observables and not to describe the dynamical aspects of the data,
no further attempt was done to optimize the input parameters of the code.
The standard built-in parameters \cite{LOP92} were used, 
especially a density for the simultaneous breakup szenario of one half of normal
nuclear density which is much larger than the values extracted from statistical
multifragmentation models (~$\rho/ \rho_0$~=~0.3 in the Copenhagen and the 
Moscow code 
and 0.135 in the MCFRAG code~).

%\noindent
If the sensitivity of the chosen observables is to be tested, it is 
--~however~-- important that the simulations provide a sample of 
Monte Carlo data which matches -- with respect to the fragment composition~--
the experimental data. This is demonstrated in figure \ref{z123}, where 
for both breakup scenarios the mean charges $\langle Z_i \rangle$ 
in ternary events, ordered
according to their sizes, are compared to the experimental data. 
The large fluctuations for the simultaneous breakup scenario are due ro the 
fact that 
only very few events with small $\Omega_{\Delta}$ values are produced (~see 
next figure~).

%\noindent
In figure \ref{omegasos}, the probability distribution of the quantity 
$\Omega_{\Delta}$ 
is shown for both breakup scenarios and the experimental data. As a reference, 
the $\Omega_{\Delta}$-distribution for a thermal system containing three
non-interacting fragments is included. For the simultaneous breakup,
the probability of stretched velocity configurations -- i.e.~ 
small $\Omega_{\Delta}$ -- is significantly smaller than for a purely
sequential decay process and for the limit of a thermal system. This 
difference was to be 
expected, since the repulsive mutual Coulomb interaction shifts initially 
stretched 
velocity configurations to larger values of $\Omega_{\Delta}$. The influence 
of the
Coulomb interaction is especially strong for the relatively small radius used in
this simulation, which is already an indication that smaller densities will 
lead to a better description of the experimental data.
Only due to this repulsion, the velocity configuration
is an image of the breakup configuration in the coordinate space. Any thermal 
motion
-- i.e.~any motion which is independent of the relative positions of the 
fragments --
reduces this correlation. For realistic input parameters of the decaying system 
(see section \ref{SEC4}), the 
correlation coefficient $r(\Omega_{\Delta}, X_{\Delta})$ between 
$\Omega_{\Delta}$ and 
the equivalent quantity in the coordinate space $X_{\Delta}$ 
\begin{equation}
  r(\Omega_{\Delta}, X_{\Delta}) = 
       \frac{\langle \Omega_{\Delta} \cdot X_{\Delta} \rangle - \langle 
             \Omega_{\Delta} \rangle
             \cdot \langle X_{\Delta} \rangle}
            {\sigma(\Omega_{\Delta}) \cdot \sigma(X_{\Delta})} 
\end{equation}
has values of approximately 0.1. (~Note that even in the case of $T=0$
and three identical charges this coefficient does not reach the value 1.0 
since the
relation between the distance of two charged particles and their relative 
momentum
due to the Coulomb-repulsion is not linear.~) If -- on the other hand -- a 
selfsimilar
radial flow dominates the momentum distribution, 
$r(\Omega_{\Delta}, X_{\Delta})$ can reach values around 0.3.

\begin{figure}
      \epsfxsize=8.0cm
      \centerline{\epsffile[0 0 550 510]{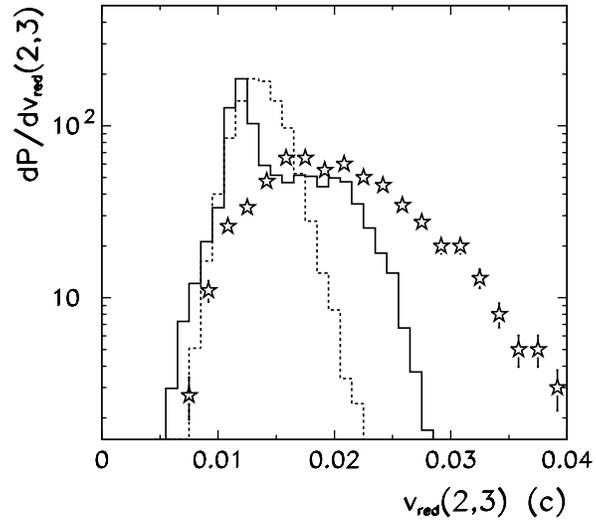}}
 \caption [] {\protect
  Probability distribution of the reduced relative velocity $v_{red}(2,3)$
  between the second and third largest fragment. The symbols refer to the 
  combined data of all targets.
  Predictions of the SOS-code assuming a sequential 
  or prompt break-up are shown by the solid and dashed histograms,
  respectively.
 }
 \label{pvrel23sos}
 \vspace{0.1cm}
 \end{figure}

%\noindent
In figure \ref{pvrel23sos}, the probability distribution of the reduced 
relative velocity $v_{red}(2,3)$ between the second and third largest 
fragments is shown, again both for the data and the SOS calculations.
The two scenarios predict significantly different relative
velocity distributions which in both cases differ clearly from the data.
In particular, the sequential calculations (solid histogram)
show a pronounced peak at $v_{red}(2,3) =0.012~c$.
This structure originates from the direct splitting of an intermediate
state into the observed fragments 2 and 3 at a rather late stage of the 
decay sequence.
The absence of this structure in the data may therefore signal either 
a smearing of the relative velocity between the final fragments 2 and 3 by
decays following the splitting into the primordial second and third largest 
fragments, or proximity effects caused by the
presence of other particles, or a different decay mechanism which does not 
produce fragment 2 and 3 via a binary splitting.

%\noindent
The results presented in figures \ref{omegasos} and \ref{pvrel23sos} suggest 
that the
quantities chosen to describe the dynamics of the multifragment events are 
sensitive
to important characteristics of the decay process. In the following chapter, 
the experimental results
will be compared to calculations with statistical multifagmentation models 
and classical three-body calculations in order
to limit the parameter space of the break-up scenario.

\section{Comparison to model calculations}
\label{SEC4}

\subsection{Statistical multifragmentation models}
\label{SEC41}

%\noindent
Since statistical multifragmentation models have been shown to describe 
the observables in the partition space of the multifragmentation process 
\cite{KRE93,BAO93,BAR93,BOT92,BOT95a,XI97}, it is the obvious next step 
to compare their predictions to the kinetic energy distribution obtained
in the present experiment. (~It should be emphasized that the description
of the partition space comprises the cross sections for binary and 
ternary events as defined in \ref{SEC31}.~)
Results are shown for the Berlin code (MCFRAG) as well as the Copenhagen
and the Moscow code. A detailed description of the differences between the 
three 
models can be found in reference \cite{GRO94}. An extensive and detailed 
investigation of 
all dynamical observables
as defined in sections \ref{SEC31} and \ref{SEC32} was only performed using
the statistical multifragmentation code MCFRAG  \cite{BAO93}.

\begin{figure}
      \epsfxsize=8.0cm
      \centerline{\epsffile[40 10 460 560]{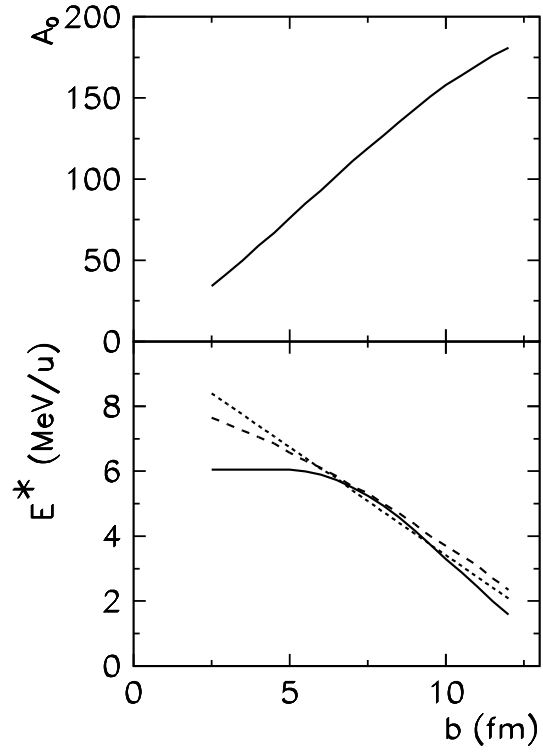}}
 \caption [] {\protect
  Input parameters for the simulations with statistical multifragmentation 
  codes.\\
  (top): Size of the decaying spectator $A_0$ versus the
   impact parameter $b$. \\
  (bottom): Excitation energy $E^*$ per nucleon versus the impact parameter.
   The short dashed, long dashed and solid lines show the values used for 
   the three codes COPENHAGEN, MOSCOW and BERLIN, respectively. \\
 }
 \label{input}
 \vspace{0.1cm}
 \end{figure}

%\noindent
All three models assume an equilibrated
source with a given number of nucleons $A$ at a density $\rho$
with an excitation energy $E^*$ per nucleon. This source is non-homogeneous, it
consists of regions of liquid with normal nuclear density and regions of gas.
To compare the 
calculations to the experimentally observed decay of the projectile spectator, 
the global parameters $A$ and $E^*$ have to be provided as a function of the 
impact parameter~$b$. To do so, 
the number of nucleons of the projectile 
spectator was calculated within a geometrical abrasion picture
for the collision Au+Au using a radius parameter of 1.3~fm. The excitation 
energy for a given spectator size within the three codes was then 
chosen according to \cite{BAO93,BAR93,BOT92}. 
For the nuclear density at freeze out the standard values of the models were 
taken,
i.e.~$\rho/ \rho_0$~=~0.3 for the Copenhagen and the Moscow code and 0.135 
for the MCFRAG code.  
In figure \ref{input}, the size of the projectile spectator and its excitation
energy are shown versus the impact parameter. It should be noted that for all 
three models
the excitation energy necessary to describe the partition space of the 
multifragmentation
is significantly smaller than the experimental results obtained for the 
reaction Au~+~Au
at 600~MeV/nucleon using a total energy balance \cite{POC95}.
The number of events to be produced for each
interval in $b$ was chosen according to the geometrical cross section for
the interval, $dP(b) \sim bdb$. The impact parameter was varied between 0.5 
and 12.0~fm in
steps of 0.5~fm. For the MCFRAG code, the calculation of the observables was 
done twice:
First, the output of the simulations was used directly, then 
random errors on the order of the experimental uncertainties for light 
particles were added to the masses and 
momenta of the fragments before the same analysis was performed.
In this way, an upper estimation for the uncertainties produced by the 
experimental resolution was achieved.

\begin{figure}
      \epsfxsize=8.0cm
      \centerline{\epsffile[20 40 560 530]{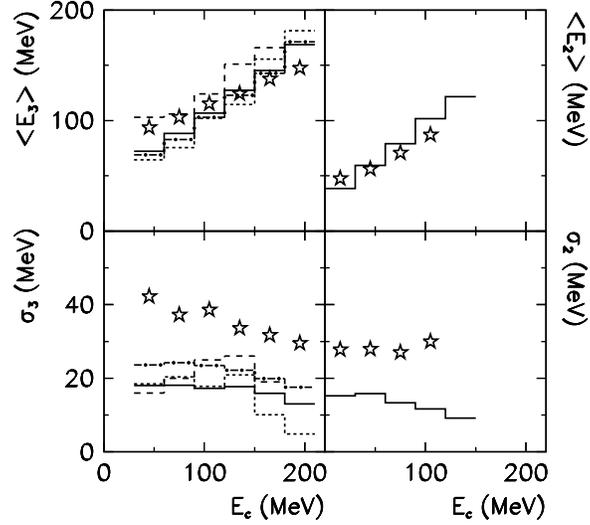}}
 \caption [] {\protect
  Mean kinetic energies $\langle E_3\rangle$ and $\langle E_2\rangle$ (top) 
  and the standard deviations $\sigma_3$ and $\sigma_2$ (bottom) as a 
  function of the nominal Coulomb-energy $E_c$. 
  The symbols denote the experimental data averaged over the reactions Au+C, Al,
  Cu and Pb. The short dashed, long dashed and solid histograms present 
  predictions of the
  COPENHAGEN, MOSCOW and MCFRAG fragmentation models, respectively. 
  The dashed-dotted
  line shows the result of a classical trajectory calculation as described in 
  section \ref{SEC42}. All calculations underpredict significantly the width of
  the energy distribution while the mean value of the kinetic energy is well 
  described.
 }
 \label{e3_ec}
 \vspace{0.1cm}
 \end{figure}

%\noindent
In figure \ref{e3_ec} the mean kinetic energies $\langle E_2 \rangle$ and 
$\langle E_3 \rangle$ for binary and ternary events 
(~as defined in \ref{SEC31}~) in the center of mass frame of the two or 
three particles  and the widths of these distributions 
$\sigma_2$ and $\sigma_3$ are plotted versus the nominal 
Coulomb energy. In the case of the MCFRAG code, the results including the 
experimental
resolution are shown, for the two other sets of simulations, the 
uncertainties due to the
experimental errors were added quadratically to the intrinsic widths of 
the energy distributions.
The mean kinetic energy $\langle E_3 \rangle$ is reasonably well 
described by all models, 
although small differences arise: For the whole range of $E_{c}$, the 
calculations using the Copenhagen model is steeper than the experimental 
distribution,
therefore the agreement is, compared to the two other models, worse. The overall
agreement of data and the three sets of calculations --~independent of 
internal details in the
theoretical treatment of the fragmentation process~-- and the simultaneous 
description 
of $\langle E_2 \rangle$ and $\langle E_3 \rangle$ by the MCFRAG code
are nevertheless a confirmation for the 
expansion of the nuclear matter prior to its decay.
The width of the energy distributions, on the other hand, is underestimated 
by almost a factor
of two both for events with two and three heavy fragments in the exit channel. 
In spite of deviations between the three sets of calculations,
the inadequate
description of $\sigma_3$ is a generic problem of all three
statistical multifragmentation models. 
Using the MCFRAG code, it was verified that this underprediction of
$\sigma_3$ can not be compensated by reasonable fluctuations of the
initial excitation energy of a given spectator: Combining the events from 
three sets of
calculations  with 0.9, 1.0 and 1.1 times $E^*(A_0)$ does not change the 
width of the energy distribution. This variation of the excitation energy 
corresponds
within the relevant range of spectator sizes approximately to the width of 
the energy 
distribution used in \cite{BOT95a} to describe the experimental charge 
distributions.

%\noindent
In figure \ref{omegastat} the probability distribution for the quantity 
$\Omega_{\Delta}$ is plotted both for the data and the calculations with the
MCFRAG code. The
calculated distribution is significantly steeper than the experimental one.
On the other hand, it is less steep than the result of the SOS-calculation 
for a 
simultaneous breakup presented in figure \ref{omegasos}. Since in both cases
the excitation energy transferred to the 
spectator matter of a given size is identical and the breakup pattern is on 
average very
similar, any differences in the velocity distributions are caused by the
different radii of the breakup volume. This will result in different 
contributions from the
Coulomb interaction and   --~more important --~in different spacial 
breakup configurations.
On average, an elongated structure will result in a smaller value of 
$\Omega_{\Delta}$
than a more compact one. If, however, the volume is very small like in 
the case of the
SOS-calculations, elongated configurations are less likely. The probability
distribution of $\Omega_{\Delta}$ is therefore expected to be steeper 
than for the more 
dilute system used for the MCFRAG-calculations.

\subsection{The influence of angular momentum}
\label{SEC43}

%\noindent
The simulations presented in the previous section showed that the
experimental energy distributions can not be explained 
in a purely thermal description of the nuclear matter, if the 
temperature is adjusted to reproduce the charge distributions.
It was shown earlier that the coupling of random and collective
motion increases the fluctuations of the kinetic energy \cite{MIL93}.
As an additional degree of freedom angular momentum was therefore 
taken into account. It is well known from the study of fission and compound
nuclei at lower energies that in heavy ion reactions
very large angular momenta can be transferred, causing a collective
rotation of the excited matter. 
INC-calculations at 100 and 200~MeV/nucleon show that the mean angular
momentum per nucleon transferred can be as large as 0.75$\hbar$, but even more
important than the mean values are the huge angular momentum fluctuations which
may reach 0.5$\hbar$ per nucleon FWHM \cite{BLA92}. 
The influence of angular momentum on the decay pattern of nuclear matter
within the framework of statistical fragmentation models has only barely 
been studied so far.

%\noindent
Calculations with the MCFRAG model were done using a version of 
the code where the treatment of angular momentum was implemented in a fully 
microcanonical way \cite{BOT95}. The impact parameter was again varied between
3.0 and 12.0~fm in steps of 0.5~fm (~below 5~fm, no events with three heavy
fragments are produced~)
and a total number of 570000 events for each set of simulations was produced.
In this implementation, the rotational degrees of freedom are assumed
to be completely thermalized and the contribution of the intrinsic
rotation of the produced fragments to the total angular momentum is neglected.
This is supposed to be a good approximation for expanded systems at the 
time of freeze out, since
the main part of the angular momentum is contained in the orbital motion 
of the fragments
around the common center of mass.

\begin{figure}
      \epsfxsize=8.0cm
      \centerline{\epsffile[0 50 490 530]{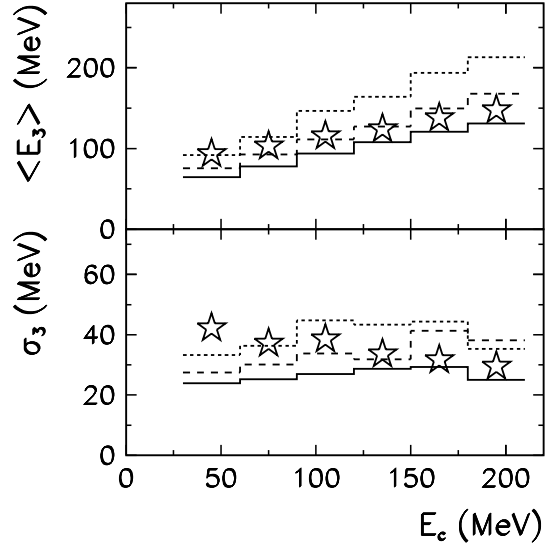}}
 \caption [] {\protect
  Mean kinetic energy $\langle E_3 \rangle$ and the standard deviation 
  $\sigma_3$
  as a function of the nominal Coulomb energy $E_c$ under the assumption of a 
  mean
  angular momentum transfer to the spectator of 0.75$\hbar$/nucleon. 
  The dotted, dashed and solid histograms
  present MCFRAG-calculations with freeze out densities of 0.055$\rho_0$, 
  0.080$\rho_0$ and 0.135$\rho_0$, respectively.
  The symbols represent the experimental data. 
 }
 \label{ekin3ang1}
 \vspace{0.1cm}
 \end{figure}

\begin{figure}
      \epsfxsize=8.0cm
      \centerline{\epsffile[0 0 550 510]{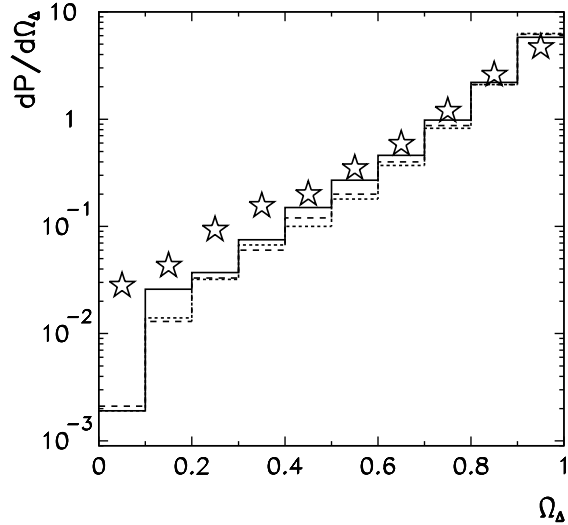}}
 \caption [] {\protect
  Probability distribution for $\Omega_{\Delta}$ under the assumption of a mean
  angular momentum transfer to the spectator of 0.75$\hbar$/nucleon. 
  The dotted, dashed and solid histograms
  present MCFRAG-calculations with freeze out densities of 
  0.055$\rho_0$, 0.080$\rho_0$ and 0.135$\rho_0$, respectively.
  The symbols denote the experimental data.
 }
 \label{omegaang1}
 \vspace{0.1cm}
 \end{figure}

%\noindent
Calculations were performed for three nuclear densities 
0.055$\rho_0$, 0.080$\rho_0$, 0.135$\rho_0$,
using the relations between impact 
parameter, system size and excitation energy which were already shown in 
figure \ref{input},
and a mean angular momentum $\langle L \rangle$ of 0.75$\hbar$A.
The angular momentum transfer was distributed according to 
\begin{equation}
  P(L) = \frac{L}{0.5 \cdot \langle L \rangle} \cdot
  \exp \left( \frac{-L}{0.5 \cdot \langle L \rangle} \right) \quad .
\end{equation}
In reference \cite{BOT95}, it was already shown that simulations with 
this angular momentum distribution together with a 
nuclear density of 0.08$\rho_0$ describe simultaneously 
the quantities $\langle E_3 \rangle$ and $\sigma_3$. The results, again 
including the influence
of the experimental uncertainties, are
shown in figure \ref{ekin3ang1} for the three densities listed above.
As expected, the mean kinetic energy as well as the width of the energy 
distribution 
increases with increasing nuclear density. Due to the fact 
that the mean rotational
energy is not very large, the incorporation of angular momentum does not change 
$\langle E_3 \rangle$ very much, as a comparison to figure \ref{e3_ec} 
demonstrates,
but the large variation of angular momenta produces nonthermal fluctuations 
which increase the
value of $\sigma_3$ significantly, resulting in a good description of both 
$\langle E_3 \rangle$
and $\sigma_3$ for densities between 0.055$\rho_0$ and 0.080$\rho_0$.
At the same time, the quantity $\Omega_{\Delta}$ is much better described, 
as is shown in figure \ref{omegaang1} where the probability distribution of 
$\Omega_{\Delta}$ is plotted for the three densities.
Independent of the nuclear density chosen the probability for the occurence 
of stretched configurations of the three velocity vectors is enhanced.

\begin{figure}
      \epsfxsize=8.0cm
      \centerline{\epsffile[0 20 570 540]{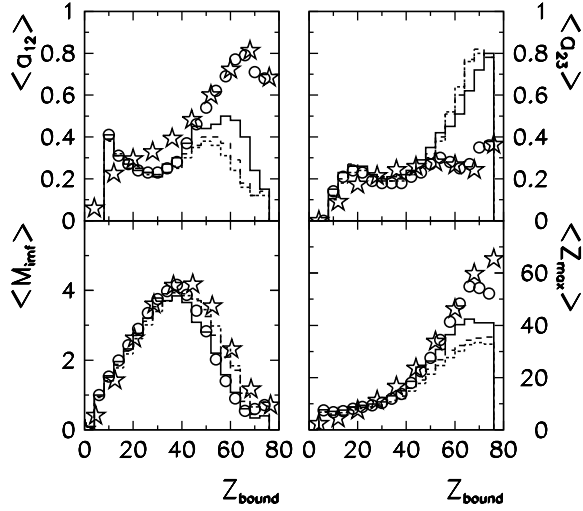}}
 \caption [] {\protect
  Mean charge asymmetry between the two largest fragments $\langle a_{12} 
  \rangle$ 
  (top left) and between the second and third largest fragment 
  $\langle a_{23} \rangle$ (top right), 
  mean number of intermediate mass fragments $\langle M_{imf} \rangle$ 
  (bottom left) and 
  mean charge of the largest fragment $\langle Z_{max} \rangle $ 
  (bottom right) versus $Z_{bound}$ under the assumption of a mean
  angular momentum transfer to the spectator of 0.75$\hbar$/nucleon.
  The stars denote the experimental data. The dotted, dashed and solid 
  histograms
  present calculations with freeze out densities of 0.055$\rho_0$, 0.080$\rho_0$
  and 0.135$\rho_0$, respectively. For comparison, the results of calculations
  without angluar momentum transfer and a freeze out density of 0.135$\rho_0$ 
  are included as open circles.
 }
 \label{partition1}
 \vspace{0.1cm}
 \end{figure}

%\noindent
To check whether the decay pattern is changed by the angular momentum, 
observables 
which were used in earlier papers \cite{KRE93,SCH96} to describe
the charge partition space of the reaction were investigated:
The mean values of the asymmetries 
\begin{equation}
a_{12} =\frac{Z_1 - Z_2}{Z_1 + Z_2} \qquad \mbox{and} \qquad 
 a_{23} =\frac{Z_2 - Z_3}{Z_2 + Z_3} \quad ,
\end{equation}

the mean number of intermediate mass fragments $M_{imf}$ and the average 
charge of the
largest fragment $Z_{max}$ are calculated as a function of $Z_{bound}$. In 
figure
\ref{partition1}, the results are shown for simulations with and without 
angular momentum
together with the experimental data. Whereas the mean number of intermediate 
mass
fragments $\langle M_{imf}\rangle$ does not change very much under the 
influence 
of angular momentum, this is not true for the details of the decay pattern of 
the spectator: The mean asymmetry 
$\langle a_{12}\rangle$ between the
charges of the largest and the second largest fragment decreases 
dramatically for 
values of $Z_{bound}$ above 50, which means that the two fragments become more
comparable in size. As a consequence, the mean charge of the largest 
fragment $\langle Z_{max}\rangle$ within an
event also decreases. At the same time, the mean asymmetry between the 
charges of the 
second and the third largest fragment $\langle a_{23}\rangle$ increases, 
which means that in the presence of angular momentum the charge of the spectator
is more evenly divided between the two
largest fragments. The changes in the breakup pattern are more pronounced for 
a small freeze out density. These results are in qualitative agreement with the
investigations presented by Botvina and Gross \cite{BOT95}, where the 
size of the largest 
fragment and the relative size of the two largest fragments was studied 
under the influence of different amounts of angular momentum.

%\noindent
From the calculations presented above it is obvious that large angular momenta 
per nucleon destroy the agreement between the results of the statistical 
multifragmentation code and the data as far as the partition pattern of the
spectator matter is concerned. This is especially true for large values of
$Z_{bound}$, i.e.~for peripheral collisions. On the other hand, it was shown
that the additional degree of freedom increased the fluctuations of the kinetic
energy by a substantial amount. The question therefore arises whether a better
overall agreement can be achieved if the transfer of angular momentum per 
nucleon to the system is reduced for large impact parameters.

%\noindent
If the peripheral reactions are treated in the abrasion-ablation picture 
applying
the formalism described in \cite{GAI91}, values for the angular momentum 
transfer 
are obtained which are smaller than the value of 0.75$\hbar$/nucleon by a factor
of 5 to 10. These numbers together with a density of 0.135$\rho_0$ 
result in a reasonable description of the partition but the energy fluctuations
are again underestimated. The mean values of the asymmetries $a_{12}$ and 
$a_{23}$
might suggest that this can be 
compensated by an increase of the nuclear density at breakup. Unfortunately, 
this
is in contradiction to the description of the quantity $\langle M_{imf}\rangle$.
The probability to find large 
values of $\langle M_{imf}\rangle$ for the $Z_{bound}$ range between 40 and 70 
decreases with increasing density. As the mean multiplicity of IMF's   
is already too small, a further 
increase of the density would make the deviations even worse.

%\noindent
This leaves no room for a parametrization of angular momentum transfer and 
density 
which fits both aspects of the experimental 
data. The charge partition space and the dynamics of multifragmentation 
events can not be described simultaneously by the statistical multifragmentation
model even if angular momentum as an additional degree of freedom and therefore 
as a potential source for fluctuations is taken into account.

%\noindent
The conclusions drawn in this section are valid only for a nuclear system where
all degrees of freedom are completely thermalized. If this is not true, i.e.~if 
the time scale for the equilibration of the rotational degrees of freedom is 
large compared
to that of the thermalization of the excitation energy, the process of 
fragmentation is decoupled from the angular momentum transfer. 
In this case, the amount of angular momentum transferred to the spectator
does not influence the partition space of the reaction, it only contributes
to the final momentum distribution of the fragments. Therefore, density and 
excitation energy on
the one hand and angular momentum on the other hand can be adjusted 
independently
and a reasonable agreement with the experimental  data can be achieved. 
This approach 
has been adopted by the Multics/Miniball group \cite{DAG97}.
It has to be stated, though, that with this modification the fragmentation 
process
is not treated in a purely microcanonical picture any longer.

\subsection{Classical three-body calculations}
\label{SEC42}

A collective
radial motion of all constituents of the spectator is another conveivable
source of fluctuations of the kinetic energy.
If the nuclear matter is compressed in the initial stage of the
reaction, an additional nonequilibrated  collective contribution to the 
motion of
the nuclear matter will be present \cite{HOF76,POG95}. Even though this 
effect is expected to be 
small in the peripheral collisions discussed in this paper, values for the 
radial flow
energy up to 1.5 MeV can not be ruled out \cite{SCH96}. First attempts have 
been made to
include collective radial flow in statistical models \cite{SUB96},
but a consistent implemention is not yet available. Therefore,
classical three-body calculations were performed to get a quantitative 
estimate for  the influence of collective flow.

%\noindent
The simultaneous emission out of a given volume is 
modeled in the following way: The centers of three non-overlapping fragments 
with a radius 
of $ 1.2 \cdot A^{1/3}$ are
distributed randomly within a sphere of radius~$R$. To each fragment, an 
isotropically
distributed initial velocity is assigned. Constrained by momentum conservation, 
these velocities were selected according to a probability distribution for the
relative kinetic energy 
\begin{equation}
  P(E) \sim E^{\alpha} \cdot \exp \left( - \frac{E}{T} \right) 
\end{equation}
 with $\alpha$ equals
0.5 or 1.0, corresponding to a volume or a surface emission of the fragments. 
In addition to
this random motion, an initial radial flow velocity 
\begin{equation}
   \vec{v}_{f,i} = \sqrt{ \frac{ 2 \cdot \epsilon_f}{m_0}} \cdot 
\frac{\vec{d}_{i}}{R}
\end{equation}
was added to the random velocities of the thermal motion. Here, 
$\vec{d}_{i}$ is the 
position of the center of
fragment $i$ with respect to the center of mass, $\epsilon_f$ is the
flow energy per nucleon for fragments located at $d_{i}$ = $R$. The charges and
masses of the fragments were obtained by a Monte Carlo sampling of the 
experimental events, thus reducing significantly the uncertainties associated
with the fragment distribution. In order to account for the recoil from 
light particles emitted sequentially from the initial fragments 
($Z_{i}^{\prime},A_{i}^{\prime}$), the measured
charges $Z_i$ and masses $A_i$ were multiplied by a factor 
$(1-T^2/(a \cdot \Delta))^{-1}$. For this correction, a level density parameter 
of $a$=10~MeV was used. The quantity $\Delta$ represents the average energy
removed by the emission of a nucleon. For simplicity, 
$ \Delta = 2T+E_s+E_b$ was assumed, where $E_s$=8~MeV and $E_b$=4~MeV 
are the typical separation energy and barrier height, respectively.
After the interaction of the primordial fragments 
($Z_{i}^{\prime},A_{i}^{\prime}$) 
has ceased, the
sequential emission of light particles leading to the
observed masses and charges ($Z_{i},A_{i}$) was assumed to take place.
For each event, the temperature parameter~$T$
was chosen according to the experimental value of $Z_{bound}$ from the relation
\begin{equation}
  T = f_T \cdot \sqrt{2\cdot(79-Z_{bound})} \quad , 
\end{equation}
where $f_T$ is a free parameter. For
$f_T$ = 1, the relation describes within the relevant range of $Z_{bound}$
reasonably well
the temperatures of the initial projectile spectators as predicted by 
microscopic transport calculations \cite{BAU88,HUB92,KRE93}. A value of 0.75
is in agreement with experimental results obtained by the He-Li isotope
thermometer \cite{POC95,XI97}.
The paths of 
the fragments were calculated under the influence of their
mutual Coulomb field and two-fragment proximity forces
according to Ref. \cite{LOP89}. Since for the further analysis those
trajectories were rejected for which the fragments overlapped during
the propagation, the influence of the proximity force turned out to be 
rather small.

\begin{figure}
      \epsfxsize=8.0cm
     \centerline{\epsffile[0 0 550 510]{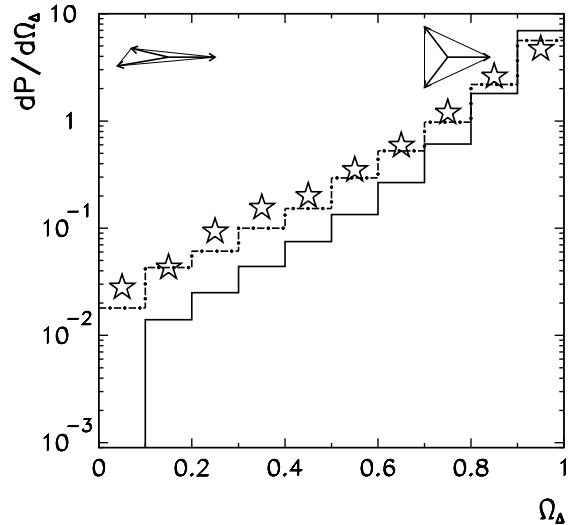}} 
 \caption [] {\protect
  Probability distribution for $\Omega_{\Delta}$. The symbols denote the
  experimental data, the histograms model calculations: 
  The solid line represents the calculation with the MCFRAG code using the 
  standard input parameters of the model. The dashed-dotted
  line shows the result of a classical trajectory calculation as described in 
  section \ref{SEC42}, where the same input parameters were used.
 }
 \label{omegastat}
 \vspace{0.1cm}
 \end{figure}

%\noindent
In a first step, these schematic trajectory calculations were 
performed with input parameters corresponding on average to those of 
the statistical model MCFRAG,
i.e.~$\alpha = 0.5$, $f_T \approx$ 0.6 - 0.8, $R \approx$ 7 - 9 fm and 
$\epsilon_f = 0$.
The results for $\langle E_3 \rangle$ and $\sigma_3$ are
comparable to those of the statistical model calculations, especially 
the width $\sigma_3$ is again significantly underpredicted in this case. 
In order
to demonstrate this, the schematic calculations for $f_T$ = 0.7, $R$ = 8~fm
and $\epsilon_f$ = 0 are included in figure \ref{e3_ec}.
The agreement of the classical calculations and the statistical model
calculations for a similar set of external parameters is a consistency check 
and shows in addition that 
the neglection of the influence of the lighter particles produced
in the reaction i.e.~the restriction of the experimental investigation to the 
two or three heaviest fragments does not change the results significantly.
The probability distribution of the quantity $\Omega_{\Delta}$ is also compared
to the results of the statistical model calculation (~see figure 
\ref{omegastat}~).

\begin{figure}
      \epsfxsize=8.0cm
      \centerline{\epsffile[0 70 560 430]{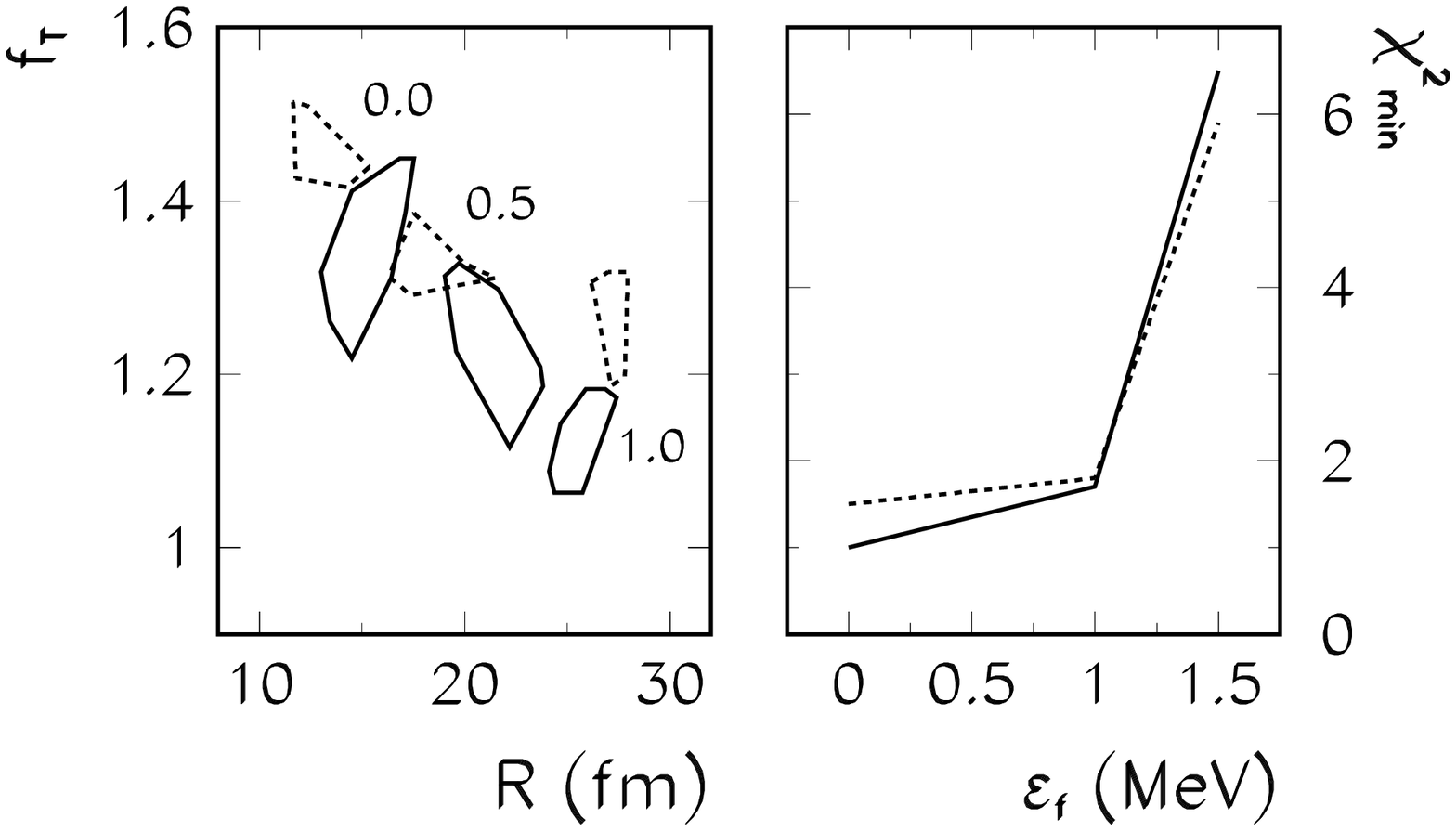}}
 \caption [] {\protect
  Parameter adjustment for the classical three-body calculations.
  The solid lines correspond to surface emission ($\alpha=1.0$), whereas the 
  dashed lines
  show the results for volume emission ($\alpha=0.5$).\\
  (left): Contour lines for $\chi^2$=2 in a plane defined by the volume 
  radius R and
  the scaling factor $f_T$ of the temperature for flow parameters 
  $\epsilon_f$= 0.0, 0.5 and 1.0 MeV. \\
  (right): Minima of the $\chi^2$-distribution as a function of $\epsilon_f$.
  Values of $\epsilon_f$ larger than 1~MeV are ruled out whereas values between 
  0 and 1~MeV show no significant differences in $\chi^2_{min}$.
 }
 \label{classical3}
 \vspace{0.1cm}
 \end{figure}

\begin{figure}
      \epsfxsize=8.0cm
      \centerline{\epsffile[0 0 570 570]{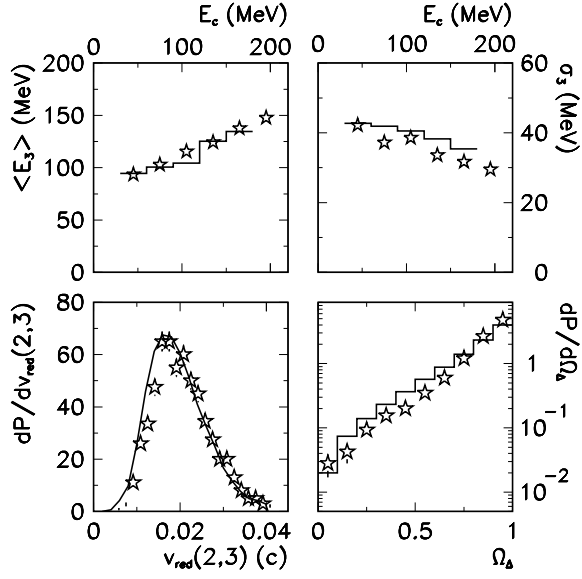}}
 \caption [] {\protect
  Mean kinetic energy $\langle E_3 \rangle$ and its standard deviation 
  $\sigma_3$,
  $v_{red}(2,3)$ and $\Omega_{\Delta}$ for classical three-body calculations
  (histograms). The parameter 
  set $R$=22~fm, $\epsilon_f$= 0.5 and $f_T$=1.2, corresponding to the 
  minimum in
  $\chi ^2$ in the case of surface emission of the three fragments 
  (~$\alpha=1.0$~) was 
  chosen. The symbols represent the experimental data.
 }
 \label{classic}
 \vspace{0.1cm}
 \end{figure}

%\noindent
In a next step, the quantities $R$, $\epsilon_f$ and $f_T$ were varied to 
fit the
experimental data.
In order to quantify the agreement between the simulations and the
experimental observations, a reduced $\chi^2$ was calculated for
each parameter set:
\begin{equation}
  \chi^2 = \frac{1}{5} \sum_{i=1}^5 \frac{ (\omega_i - \mu_i)^2}
                       { \delta_i^2} \quad .
  \label{EQ_6}
\end{equation}
Here, $\omega_i$ are the four coefficients characterizing the
fits to the three-particle data in figure \ref{ekin3}
and, in addition, the
mean reduced velocity
between the two lighter fragments as shown in figure \ref{pvrel23sos}.
$\delta_i$ and $\mu_i$ denote the experimental uncertainties of these
quantities and the corresponding model predictions, respectively.
The result is shown in figure \ref{classical3}. A clear minimum of
$\chi^2$ can be determined for each given flow parameter
$\epsilon_f$ by varying independently the other two model parameters
$R$ and $f_T$. The left part of figure \ref{classical3} shows in
a $R$ - $f_T$ plane the contour lines with $\chi^2$ = 2
for $\epsilon_f$ = 0 ($R \approx$ 15 fm), 0.5
($R \approx$ 22 fm) and 1 MeV ($R \approx$ 26 fm)
and for the two values of  the exponent $\alpha$.
The corresponding minima of the $\chi^2$-distribution are displayed 
in the right panel of figure \ref{classical3}
as a function of $\epsilon_f$. Both for volume emission and surface emission,
values for the flow parameter $\epsilon_f$
larger than 1 MeV are ruled out whereas the results obtained by
values between 0 and 1~MeV show no significant
difference in $\chi^2_{min}$. 
To demonstrate the quality of the parameter adjustment, the quantities 
$\langle E_3 \rangle $, $\sigma_3$ and v$_{red}(2,3)$ are shown in
figure \ref{classic} for the parameter set $R$ = 22~fm, $\epsilon_f$ = 0.5~MeV
and $f_T$ = 1.2. In the lower right part of figure \ref{classic}, 
$\Omega_{\Delta}$ --~which was
not used in the fitting procedure~-- is compared to the experimental
values. As expected from the results shown in figures
\ref{omegasos} and \ref{omegastat}, the probability for the existence
of stretched velocity configurations increases with increasing radius of 
the decaying
system. The $\Omega_{\Delta}$-distribution is nevertheless not directly 
comparable 
to those shown in figures \ref{omegasos} and \ref{omegastat}: They were 
achieved assuming
a fixed breakup density for all decaying systems, whereas the three-body 
calculations 
assume a fixed breakup volume.

%\noindent
This set of simulations suggests
the disintegration of a highly excited and rather extended
nuclear system and very low values of the flow parameter.
For $Z_{bound}$=55 -- the mean value for events with three heavy
particles in the exit channel of the reaction -- the fit values correspond
to a temperature parameter of approximately 10~MeV and a density 
below 0.05$\rho_0$ which is much smaller than the values used for the
MCFRAG calculations in order to reproduce the partition space of the
reaction.

%\noindent
In the framework of these schematic calculations
the large freeze out radius is due to the balance between Coulomb
energy and temperature:
If a higher nuclear density is assumed,  the Coulomb repulsion
is much stronger and requires therefore a compensation by a 
lower temperature parameter and a vanishing flow to describe the 
energy spectra.  The fluctuations 
$\sigma_3$ of the kinetic energy, on the other hand, 
reflect in addition to thermal fluctuations also
fluctuations due to the position sampling within the breakup volume.
Thus, lower temperatures and especially smaller radii 
lead to a significant reduction of $\sigma_3$ which can not be compensated 
by the 
small values of radial flow consistent with the energy spectra.

\section{Conclusions}
\label{SEC5}

%\noindent
Kinematic correlations between two and three heavy projectile
fragments produced in Au induced reactions at E/A = 600 MeV have been studied.
A comparison of the observables to the results of the schematic SOS-model
confirms their sensitivity to the disassembly configuration.
Classical trajectory calculations sampling the experimental charge
distribution limit significantly the possible
parameter space of the breakup scenario. Taken at their face
values these simulations require highly excited and rather extended
nuclear systems at the time of the breakup.
These source parameters differ significantly from breakup
parameters needed by statistical multifragmentation models in
order to describe
the observed fragment distributions and mean values of the kinetic
energy distributions. On the other hand, these models are not able 
to reproduce the fluctuations of the energy distribution. 
Binary events not attributed to binary fission also show 
fluctuations of the relative kinetic energy which
can only be described by the same rather high - and probably unrealistic -
thermal contribution.
The introduction of angular momentum into the
statistical model improves the description of the energy fluctuations,
but does not allow anymore to reproduce simultaneously the charge partition.

%\noindent
For any further attempt to reconcile the kinetic observables and the
partition pattern of the spectator matter two possible approaches
seem conceivable: Either the assumption of a global equilibrium
established prior to the fragmentation process is oversimplified and 
has to be given up, or the statistical models have to be refined.

%\noindent
The nuclear interaction during the breakup process, for example, is
so far ignored, the interaction between the fragments is limited to the 
Coulomb repulsion.
( In the classical three-body trajectory calculations present in this work,
a nuclear proximity potential is included, but its influence is strongly
suppressed by the requirement that the fragments do not overlap.~)
One might speculate that in the case of a stronger overlap of the
fragments in an earlier stage of the breakup, the nuclear
attractive force between the fragments may partially compensate
the Coulomb repulsion. Thus, smaller radii would not
necessarily lead to an overestimation of the kinetic energies.
First steps to add the nuclear interaction
between the fragments to statistical decay models
in a consistent manner have already been undertaken \cite{SAT90,DAS93}. A recent
publication suggests that the nuclear interaction is indeed relevant
for excitation energies up to approximately 10~MeV/u \cite{DAS97}.
At the same time, large fluctuations -- similar to dissipative phenomena
and shape fluctuations known to be important in binary
fission \cite{GOE91} -- may arise.

%\noindent
On the long term, a quantitative understanding of
fluctuations and their development during the disassembly phase clearly
requires dynamical transport models which include a realistic
treatment of fluctuations on a microscopic level.
Significant progress in the development of microscopic transport models
has been achieved during the last decade \cite{BUUQMD},
but only recently first microscopic calculations were published which
reproduce for the ALADIN data both the multiplicity of the fragments and
the slopes of their kinetic energy spectra \cite{GOS97}. In the
framework of this model -- and in line with previous studies 
\cite{BOA89,BOA89a,KUN91} --
it is found that the decaying system is not in thermal
equilibrium and that the breakup is dominated by dynamical processes.
However, the fragment composition agrees with the experimental 
one only for a short
time interval after the collision (60~fm/c) and is drastically altered
during the further time evolution. Thus, a consistent description
of the time evolution from the first stages of the collision via the formation
of primordial excited fragments to their eventual deexcitation and formation
of individual quantum states within one microscopic model is 
still not available.
First attempts to take into account the quantal nature of
the nuclear system are being pursued \cite{OHN97,SCH97,SCH97a}
for which the present data may serve as a valuable testing
ground.

%\noindent
The authors thank D.H.E.~Gross and A.S.~Botvina for
providing us with their statistical multifragmentation codes
and for helpful discussions. 
This work was partly supported by the Bundesministerium f\"ur
Forschung und Technologie.
J.P.~and M.B.~acknowledge the financial support 
of the Deutsche Forschungsgemeinschaft
under the Contract no. Po 256/2-1 and Be 1634/1-1.

\newpage

%----------------------------- figures -------------------------------------
%
%  1: setup
%  2: music
%  3: musicz
%  4: beam
%  5: masses
%  6: wqternary
%  7: multlp
%  8: betaparperp
%  9: ekin3
% 10: omega
% 11: z123
% 12: omegasos
% 13: pvrel23sos
% 14: input
% 15: e3_ec
% 16: omegastat
% 17: ekin3ang1
% 18: omegaang1
% 19: partition1
% 20: classical3
% 21: classic
%
%----------------------------- setup ---------------------------------------
%
%
%-------------------------- music -----------------------------------------
%
%
%------------------------ musicz --------------------------------------------
%
%
%----------------------- beam ----------------------------------------------
%
%
%---------------------- masses ---------------------------------------------
%
%
%---------------------- wqternary ---------------------------------------------
%
%
%---------------------- multlp ---------------------------------------------
%

\newpage		%% wfjm formatting for preprint 
%
%---------------------- betaparperp -------------------------------------------
%

\newpage		%% wfjm formatting for preprint 
%
%---------------------- ekin3 ---------------------------------------------
%
%
%---------------------- omega ---------------------------------------------
%
%
%---------------------- z123 ---------------------------------------------
%
%
%---------------------- omegasos ---------------------------------------------
%
 \begin{figure}
      \epsfxsize=8.0cm
      \centerline{\epsffile[0 0 550 510]{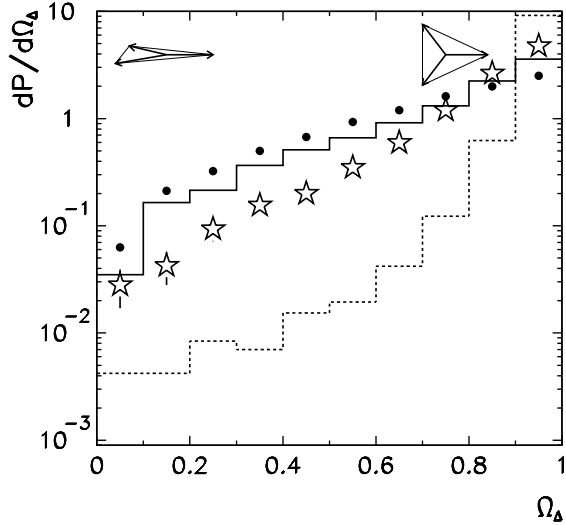}}
 \caption [] {\protect
  Probability distribution of the observable $\Omega_{\Delta}$.
  The stars refer to the combined data of all targets.
  Predictions of the SOS-code assuming a sequential 
  or prompt break-up are shown by the solid and dashed histograms,
  respectively. Also included are the values for a thermal system
  without Coulomb interaction.
  (filled circles).
 }
 \label{omegasos}
 \vspace{0.1cm}
 \end{figure}
%
%---------------------- pvrel23sos ---------------------------------------------
%

\newpage		%% wfjm formatting for preprint 
%
%---------------------- input ---------------------------------------------
%

\newpage		%% wfjm formatting for preprint 
%
%---------------------- e3_ec ---------------------------------------------
%
%
%---------------------- omegastat ---------------------------------------------
%
%
%---------------------- ekin3ang1 ---------------------------------------------
%
%
%---------------------- omegaang1 ---------------------------------------------
%
%
%---------------------- partition1 ---------------------------------------------
%
%
%---------------------- classical3 ---------------------------------------------
%
%
%---------------------- classic ------------------------------------------------
%
\end{document}